\newcommand\ckmfitter{{CKMfitter}}
\newcommand{\simgt}{\,\hbox{\lower0.6ex\hbox{$\sim$}\llap{\raise0.6ex\hbox{$>$}}}\,}
\newcommand{\simlt}{\,\hbox{\lower0.6ex\hbox{$\sim$}\llap{\raise0.6ex\hbox{$<$}}}\,} 
\newcommand{\bbq}{\ensuremath{B_q\!-\!{\bar B}_q\,}}

\newcommand{\imag}{\mathrm{Im}\,}

\newcommand{\dm}{\ensuremath{\Delta M}}
\newcommand{\dg}{\ensuremath{\Delta \Gamma}}

\newcommand\ea{{\em et al.}}

\documentclass[
  aps,
  prd,
  reprint,
  showpacs,
  groupedaddress,
  amsmath,
  amssymb,
  floatfix
]{revtex4-1}
\usepackage{graphicx,epsfig,dcolumn,multirow}
\newcolumntype{d}[1]{D{.}{.}{#1}}

\begin{document}

\title{
   Current status of the Standard Model CKM fit\\
    and constraints on $\Delta F=2$ New Physics}

\author{The \ckmfitter\  Group\\
\vspace{0.1cm}
J.~Charles$^{\,a}$,
O.~Deschamps$^{\,b}$,
S.~Descotes-Genon$^{\,c}$,
H.~Lacker$^{\,d}$,
A.~Menzel$^{\,d}$,
S.~Monteil$^{\,b}$,\\
V.~Niess$^{\,b}$,
J.~Ocariz$^{\,e}$,
J.~Orloff$^{\,b}$,
A.~Perez$^{\,f}$,
W.~Qian$^{\,g}$,
V.~Tisserand$^{\,g}$,
K.~Trabelsi$^{\,h,i}$,\\
P.~Urquijo$^{\,j}$,
L.~Vale Silva$^{\,c}$
}

\affiliation{
$^{a}$Aix Marseille Universit\'e, Universit\'e de Toulon, CNRS, CPT UMR 7332, 13288, Marseille, France\\
$^{b}$Clermont Universit\'e, Universit\'e Blaise Pascal, CNRS/IN2P3, LPC, 63170 Aubi\`ere, Clermont-Ferrand, France\\
$^{c}$Laboratoire de Physique Th\'eorique 
                   B\^{a}timent 210, Universit\'e  Paris-Sud, F-91405 Orsay Cedex, France \\
                   (UMR 8627 du CNRS  associ\'ee \`a l'Universit\'e Paris-Sud) \\
$^{d}$Humboldt-Universit\"at zu Berlin,
                   Institut f\"ur Physik,
                   Newtonstr. 15,
                   D-12489 Berlin, Germany \\
$^{e}$Laboratoire de Physique Nucl\'{e}aire et de Hautes Energies, 
IN2P3/CNRS, Universit\'{e} Pierre et Marie Curie Paris 6 et Universit\'{e} Denis
Diderot Paris 7, F-75252 Paris, France \\
 $^{f}$Institut Pluridisciplinaire Hubert Curien,
23 rue du loess - BP28, 67037 Strasbourg cedex 2, France\\
 $^{g}$Laboratoire d'Annecy-Le-Vieux de Physique des Particules 
                   9 Chemin de Bellevue, BP 110, F-74941
                   Annecy-le-Vieux Cedex, France\\
                   (UMR 5814 du CNRS-IN2P3 associ\'ee \`a
                   l'Universit\'e de Savoie)\\
$^{h}$High Energy Accelerator Research Organization, KEK 
                  1-1 Oho, Tsukuba, Ibaraki 305-0801 Japan\\
$^{i}$Ecole Polytechnique F\'ed\'erale de Lausanne (EPFL),
B\^atiment des Sciences Physiques, CH-1015 Lausanne, Switzerland\\
         $^{j}$School of Physics, University of Melbourne, Victoria 3010, Australia
}
  
\date{\today}

\begin{abstract}
This article summarises the status of the global fit of the CKM parameters within the Standard Model performed by the CKMfitter group. Special attention is paid to the inputs for the CKM angles $\alpha$ and $\gamma$ and the status of $B_s\to\mu\mu$ and $B_d\to \mu\mu$ decays. We illustrate the current situation for other unitarity triangles. We also discuss the constraints on generic $\Delta F=2$ New Physics. All results have been obtained with the \ckmfitter\ analysis package, featuring the frequentist statistical approach and using Rfit to handle theoretical uncertainties.
\end{abstract}


\maketitle

\section{Introduction}

In the Standard Model (SM), the weak charged-current transitions
mix quarks of different generations, which is encoded in the
unitary Cabibbo-Kobayashi-Maskawa (CKM) matrix~\cite{Cabibbo:1963yz,Kobayashi:1973fv}.
In the case of three generations of quarks,
the physical content of this matrix reduces to four real parameters,
among which one phase, the only source of $CP$ violation in the SM
(neglecting  $CP$-violating effects induced by the strong-interaction $\theta$-term or neutrino masses):
\begin{eqnarray}
\lambda^2&=&\frac{|V_{us}|^2}{|V_{ud}|^2+|V_{us}|^2}\,,
\quad
A^2\lambda^4 =\frac{|V_{cb}|^2}{|V_{ud}|^2+|V_{us}|^2}\,, \nonumber\\
\bar\rho+i\bar\eta&=&-\frac{V_{ud}V_{ub}^*}{V_{cd}V_{cb}^*}\,,
\end{eqnarray}
One can exploit the unitarity of the CKM matrix to determine all its elements (and when needed, to obtain their Wolfenstein expansion in powers of $\lambda$)~\cite{Wolfenstein:1983yz,Battaglia:2003in,Charles:2004jd}.

\begin{table*}
\renewcommand\arraystretch{1.2}
\caption{Constraints used for the global fit, and the main inputs involved (more information can be found in ref.~\cite{CKMfitterwebsite}). When two errors are quoted, the first one is statistical, the second one systematic. The lattice inputs are our own averages obtained as described in the text. \label{tab:expinputs}}
\begin{tabular}{c|c|cccc|ccc}
CKM  & Process  & \multicolumn{4}{c|}{Observables}  & \multicolumn{3}{c}{Theoretical inputs}\\
\hline
$|V_{ud}|$ & $0^+\to 0^+$ transitions 
                  & $|V_{ud}|_{\rm nucl}$&=& $0.97425\pm 0\pm 0.00022$
                  & \cite{TownerHardy} & \multicolumn{3}{c}{Nuclear matrix elements} \\
                    \hline
$|V_{us}|$ & $K\to\pi\ell\nu$ 
                  & $|V_{us}|_{\rm SL}f_+^{K\to\pi}(0)$&=& $ 0.21664\pm0.00048 $ & \cite{PDG}
                  & $f_+^{K\to\pi}(0)$&=& $0.9641\pm 0.0015\pm 0.045$\\
                 &  $K\to e\nu_e$ 
                 & ${\cal B}(K\to e\nu_e)$&=&$(1.581\pm0.008)\cdot 10^{-5}$ & \cite{PDG}
                 &  $f_K$&=& $155.2\pm0.2\pm0.6 $ MeV \\
                &  $K\to \mu\nu_\mu$ 
                &  ${\cal B}(K\to \mu\nu_\mu)$&=& $0.6355 \pm 0.0011$
                & \cite{PDG}\\
                 &  $\tau \to K \nu_\tau$ 
                 & ${\cal B}(\tau \to K\nu_\tau)$&=&$(0.6955 \pm 0.0096)\cdot 10^{-2}$
                 & \cite{PDG}\\
                 \hline
$\frac{|V_{us}|}{|V_{ud}|}$                 &  $K\to \mu\nu/\pi\to\mu\nu$ & 
                 $\displaystyle \frac{{\cal B}(K\to \mu\nu_\mu)}{{\cal B}(\pi \to \mu\nu_\mu)}$
                         &=&$1.3365 \pm 0.0032$
                 & \cite{PDG} &
                 $f_K/f_\pi$&=&$1.1942 \pm 0.0009\pm0.0030$ 
                  \\
                 &  $\tau\to K\nu/\tau \to \pi\nu$ &   
                 $\displaystyle \frac{{\cal B}(\tau \to K\nu_\tau)}{{\cal B}(\tau \to \pi\nu_\tau)}$
                        &=& $(6.43 \pm 0.09)\cdot 10^{-2}$
                 & \cite{PDG} \\
                 \hline
$|V_{cd}|$   & $\nu N$ & $|V_{cd}|_{\nu N}$ &=& $0.230\pm 0.011$ & \cite{PDG}\\
                   & $D\to \mu\nu $ & ${\cal B}(D\to \mu\nu)$ &=& $(3.74\pm0.17)\cdot 10^{-4}$ 
                   & \cite{HFAG}
                   &$f_{D_s}/f_D$&=&$1.201 \pm 0.004\pm0.010$\\
                   & $D\to \pi\ell\nu $ & $|V_{cd}|f_+^{D\to \pi}(0)$ &=& $0.148 \pm 0.004$ 
                   & \cite{DtopiandK}
                   &$f_+^{D\to \pi}(0)$&=&$0.666\pm 0.020\pm 0.048 $\\
                   \hline
$|V_{cs}|$   &  $W\to c\bar{s}$ &   $|V_{cs}|_{W\to c\bar{s}}$ &=& $0.94^{+0.32}_{-0.26}\pm 0.13$ & \cite{PDG}\\ 
                   & $D_s\to \tau\nu$ 
                   & ${\cal B}(D_s\to \tau\nu)$&=& $(5.55\pm0.24) \cdot 10^{-2}$ 
                   &  \cite{HFAG} 
                   & $f_{D_s}$ &=& $245.3\pm 0.5 \pm 4.5$ MeV\\
                   & $D_s\to \mu\nu$ 
                   & ${\cal B}(D_s\to \mu\nu_\mu)$&=& $(5.57\pm0.24)\cdot 10^{-3}$ 
                   &  \cite{HFAG}\\
                   & $D\to K\ell\nu $ & $|V_{cs}|f_+^{D\to K}(0)$ &=& $0.712 \pm 0.007$ 
                   & \cite{DtopiandK,Babar-DtoK}
                   &$f_+^{D\to K}(0)$&=&$ 0.747\pm0.011\pm0.034$\\
                   \hline
$|V_{ub}|$ & semileptonic decays 
                  & $|V_{ub}|_{\rm SL}$ &=& $(3.70 \pm 0.12 \pm 0.26)\cdot 10^{-3}$ 
                  &  \cite{HFAG}&
                     \multicolumn{3}{c}{form factors, shape functions}\\
                  & $B\to \tau\nu$ 
                  & ${\cal B}(B\to\tau\nu)$ &=& $(1.08\pm0.21) \cdot 10^{-4}$ & \cite{btaunu,HFAG}
                  &   $f_{B_s}/f_B$&=& $1.205\pm 0.004 \pm 0.007 $\\
                  \hline
$|V_{cb}|$ & semileptonic decays
                 & $|V_{cb}|_{\rm SL}$ &=& $(41.00\pm 0.33 \pm 0.74 )\cdot 10^{-3}$ &  \cite{HFAG}
                 &  \multicolumn{3}{c}{form factors, OPE matrix elements}\\
\hline
$\alpha$ & $B\to\pi\pi$, $\rho\pi$, $\rho\rho$ 
                & \multicolumn{3}{c}{branching ratios, $CP$ asymmetries} & \cite{HFAG} 
                & \multicolumn{3}{c}{isospin symmetry}\\
                \hline
$\beta$   & $B\to (c\bar{c}) K$ 
               & $\sin(2\beta)_{[c\bar{c}]}$ &=& $0.682 \pm 0.019$ 
               & \cite{HFAG}\\
\hline
$\gamma$ & $B\to D^{(*)} K^{(*)}$ 
                 & \multicolumn{3}{c}{inputs for the 3 methods}
                 &  \cite{HFAG}& \multicolumn{3}{c}{GGSZ, GLW, ADS methods} \\
  \hline
$\phi_s$ & $B_s\to J/\psi (KK, \pi\pi)$ & $\phi_s$ &=& $-0.015\pm 0.035$
             & \cite{HFAG}&
 \\
  \hline  
$V_{tq}^*V_{tq'}$       & $\Delta m_d$ 
                & $\Delta m_d$ &=& $0.510 \pm 0.003$ ps${}^{-1}$
                & \cite{HFAG}
                &  $\hat{B}_{B_s}/\hat{B}_{B_d}$ &=& $1.023 \pm 0.013\pm0.014$\\ 
                 & $\Delta m_s$ & $\Delta m_s$ &=& $17.757\pm0.021$ ps${}^{-1}$ 
                 & \cite{HFAG} 
                 & $\hat{B}_{B_s}$&=& $1.320\pm0.017\pm0.030$\\
                 & $B_s\to \mu\mu$ & ${\cal B}(B_s\to\mu\mu)$ &=& $(2.8^{+0.7}_{-0.6})\cdot 10^{-9}$
                 & \cite{Bsmumu} & $f_{B_s}$ &=& $225.6\pm1.1\pm5.4 $ MeV \\
                 \hline
$V_{td}^*V_{ts}$  
      & $\epsilon_K$ & $|\epsilon_K|$ &=& $(2.228\pm0.011)\cdot 10^{-3}$
       & \cite{PDG} 
       &$\hat{B}_K$&=& $0.7615\pm0.0027\pm0.0137$\\
$V_{cd}^*V_{cs}$      &   &&&&& $\kappa_\epsilon$&=& $0.940\pm0.013\pm0.023 $
\end{tabular}

\end{table*}

Extracting information on these parameters from data is a challenge for both experimentalists and theorists, since the SM depends on a large set of parameters which are not predicted within its framework, and must be determined experimentally. An additional difficulty stems from the presence of the strong interaction binding quarks into hadrons, which is responsible for most of the theoretical uncertainties discussed when determining the CKM matrix parameters. 
The CKMfitter group aims at this goal by combining a large set of constraints from flavour physics, using a standard $\chi^2$-like frequentist approach, in addition to a specific (Rfit) scheme to treat theoretical uncertainties~\cite{Charles:2004jd,CKMfitterwebsite} (see refs.~\cite{Ciuchini:2000de,Bona:2005vz,Laiho:2009eu,Eigen:2015yva} for alternative approaches in this context). 

As will be illustrated below, the SM global fit has reached a remarkable accuracy from both
the experimental and theoretical points of view. In this context, and following a long history of 
flavour as a probe for ``New Physics'' (existence of the charm quark, bounds on the top quark mass\ldots), one can also use flavour observables to constrain models of New Physics (NP), either in a particular scenario or with a rather generic scope. We will follow the second avenue, providing results for generic New Physics in $\Delta F=2$ and updating ref.~\cite{Charles:2011va}.

The results presented here correspond to the most recent update performed by the CKMfitter collaboration, including results obtained until the CKM 2014 workshop in Vienna~\cite{CKMfitterwebsite}.

\section{Inputs for the SM global fit}

\subsection{General discussion}

Not all the observables in flavour physics can be used as inputs to constrain the CKM matrix, due to limitations on our experimental and/or theoretical knowledge on these quantities. The list of inputs to the global fit is indicated in Table~\ref{tab:expinputs}: they fulfill the double requirement of a satisfying control of the attached theoretical uncertainties and a good experimental accuracy of their measurements. In addition, we only take as inputs the quantities that provide constraints on the CKM parameters $A,\lambda,\bar\rho,\bar\eta$. We will see below that not all parameters are equally relevant for the global fit.

A major source of uncertainties in flavour analyses arises from matrix elements that encode the effects of the strong interaction in the nonperturbative regime, corresponding here to
decay constants, form factors and bag parameters. We rely mainly on lattice QCD simulations for the determination of these quantities, as they provide well-established methods to compute these observables with a controlled accuracy. Some of the uncertainties have a clear statistical interpretation. Lattice simulations evaluate Green functions in a Euclidean metric expressed as path integrals using Monte Carlo methods, and their accuracy depends on the size of the sample of gauge configurations used for the computation. The remaining uncertainties are  systematic: they are now dominant in most cases and they depend on the computational strategies chosen by competing lattice collaborations: discretisation methods used to describe gauge fields and fermions on a lattice, interpolating fields, parameters of the simulations, such as the size of the (finite) volumes and lattice spacings, the masses of the quarks that can be simulated, and the number of dynamical flavours included as sea quarks. These simulations must often be extrapolated to obtain physical quantities, relying in particular on effective theories such as chiral perturbation theory and heavy-quark effective theory which induce further systematics.

The combination of lattice values is a critical point of most global analyses of the flavour physics data, even though  there is no universal definition of theoretical uncertainties (and hence how to combine them). Several approaches have been proposed to perform such a combination~\cite{Laiho:2009eu,Aoki:2013ldr}, and we have
also proposed our own scheme, systematic, reproducible and to some extent conservative~\cite{Lenz:2010gu}. We have collected the relevant lattice calculations of the meson decay constants, as well as the $B_d$, $B_s$ and $K$ bag parameters, and the $K\to\pi$, $D\to \pi$ and $D\to K$ vector form factors at zero momentum transfer. We base our set of calculations on the latest FLAG (Flavour Lattice Averaging Group) report~\cite{Aoki:2013ldr}, with the addition of new results published since that report was written~\cite{CKMfitterwebsite}.  We perform our averages considering values from lattice simulations with different numbers of dynamical flavours ($N_f=2,2+1,2+1+1$). Even though the different collaborations attempt at assessing the corresponding systematics in a careful way, one cannot exclude that such combinations are affected by further systematics which unfortunately cannot be assessed easily.
These lattice averages are the input parameters used in the fits presented in this paper. In the specific case of decay constants, the $SU(3)$-flavour breaking ratios 
$f_K/f_\pi$, $f_{D_s}/f_D$, $f_{B_s}/f_{B_d}$ are better determined than the individual decay
constants. We will therefore take these ratios as well as the strange-meson decay constants
as reference quantities for our inputs. In the same spirit,
it is more relevant to consider the predictions of the ratio $K_{\ell 2}/\pi_{\ell 2}$ of the kaon and pion leptonic partial widths, as well as $\mathcal{B}(\tau\to K\nu_\tau)/\mathcal{B}(\tau\to \pi\nu_\tau)$ instead of individual branching ratios. 

There are also other sources of theoretical uncertainties. This is the case
for the inclusive and exclusive determinations of $|V_{ub}|$ and $|V_{cb}|$, which involve nonperturbative inputs of different natures. We use the latest HFAG results~\cite{HFAG} for each of these determinations and combine inclusive and exclusive determinations following the same scheme as for the combination of lattice quantities.  We also need theoretical inputs for heavy up-type quark masses, namely $\bar{m}_c(\bar{m}_c)=(1.286\pm0.013\pm0.040)$ GeV and $\bar{m}_t(\bar{m}_t)=(165.95\pm 0.35\pm 0.64)$ GeV,
as well as to the strong coupling constant $\alpha_s(M_Z)=0.1185\pm0\pm 0.0006$.
We refer the reader to refs.~\cite{
Lenz:2010gu,Lenz:2012az,Deschamps:2009rh,Charles:2013aka} for a more detailed discussion of each constraint, whereas the related hadronic inputs can be found in ref.~\cite{CKMfitterwebsite}.

\subsection{Specific inputs}

A few specific inputs have changed recently and deserve comment.

Constraints on the CKM angle $\alpha$ are derived from the isospin analysis of the charmless $B^{\pm,0}\to(\pi\pi)^{\pm,0}$, $B^{\pm,0}\to(\rho\rho)^{\pm,0}$ and $B^{0}\to(\rho\pi)^0$ decays. Assuming the isospin symmetry and neglecting the electroweak penguin contributions, the amplitudes of the SU(2)-conjugated modes are constrained by triangular (or pentagonal) relations. The measured branching fractions and asymmetries in the $B^{\pm,0}\to(\pi\pi)^{\pm,0}$ and $B^{\pm,0}\to(\rho\rho)^{\pm,0}$ modes and the bilinear form factors in the Dalitz analysis of the $B^{0}\to(\rho\pi)^0$ decays provide enough observables to simultaneously determine the weak phase $\beta+\gamma=\pi-\alpha$ together with the tree and penguin contributions to each mode.

\begin{figure}
\includegraphics[width=8cm]{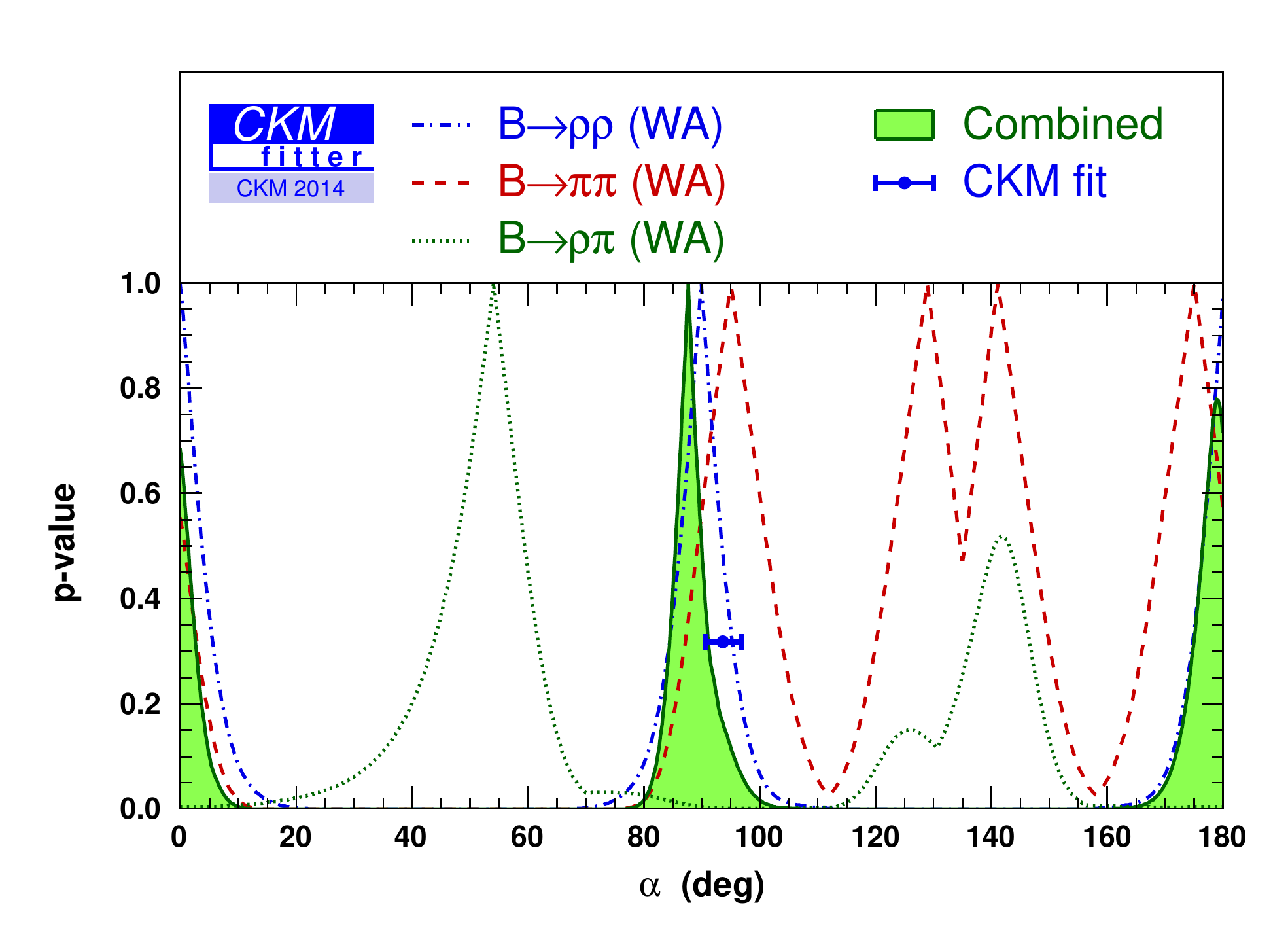}
\caption{Inputs for $\alpha$ used in the SM global fit. We show the world averages for $\pi\pi$, $\rho\pi$ and $\rho\rho$ separately.}
\label{fig:alpha}
\end{figure}

The world average constraint on $\alpha$ is so far dominated by the $B^{\pm,0}\to(\rho\rho)^{\pm,0}$ data, thanks to the low level of the penguin contribution to these modes, conducting to the 68.3\% Confidence Level ($CL$) intervals :
\begin{equation}
\alpha(B\to\rho\rho) = (89.9^{+5.4}_{-5.3})^{\circ} \cup (0.1^{+5.3}_{-5.4})^{\circ}.
\end{equation}
The recent update of the measured branching fraction of the $B\to\pi^0\pi^0$ decay, driven by the Belle experiment~\cite{Vanhoefer:2014mfa}, significantly improves the determination of 
$\alpha$ through the isospin analysis of the $B^{\pm,0}\to(\pi\pi)^{\pm,0}$ modes. The 68.3\% $CL$ intervals 
\begin{equation}
\alpha(B\to\pi\pi) = (95.0^{+8.8}_{-7.9})^{\circ} \cup (175.0^{+7.9}_{-8.8})^{\circ} \cup (135.5 \pm 15)^{\circ}
\end{equation}
are obtained. Combining the experimental data for the $\pi\pi$, $\rho\rho$ and $\rho\pi$ decay modes, the world average 68.3\% $CL$ intervals 
\begin{equation}
\alpha_{WA} = (87.7^{+3.5}_{-3.3})^{\circ} \cup (179.0^{+3.7}_{-4.1})^{\circ} 
\end{equation}
are obtained (Fig.~\ref{fig:alpha}). The recent Belle update on ${\cal B}(B^0\to\pi^0\pi^0)$ improves the 1$\sigma$ $\alpha$ resolution by $0.5^{\circ}$ with respect to the previous determination.

\begin{figure}
\includegraphics[width=8cm]{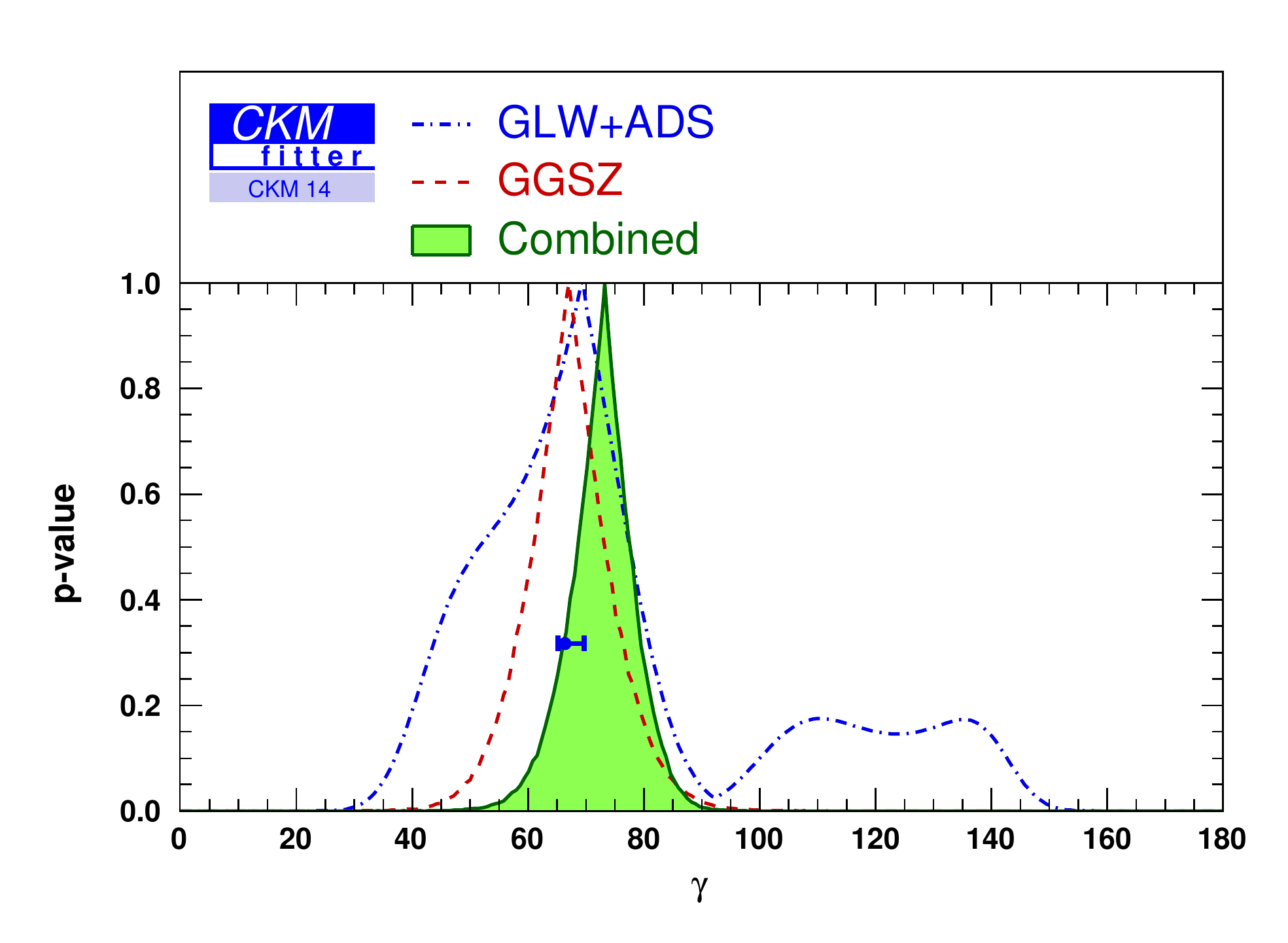}
\includegraphics[width=8cm]{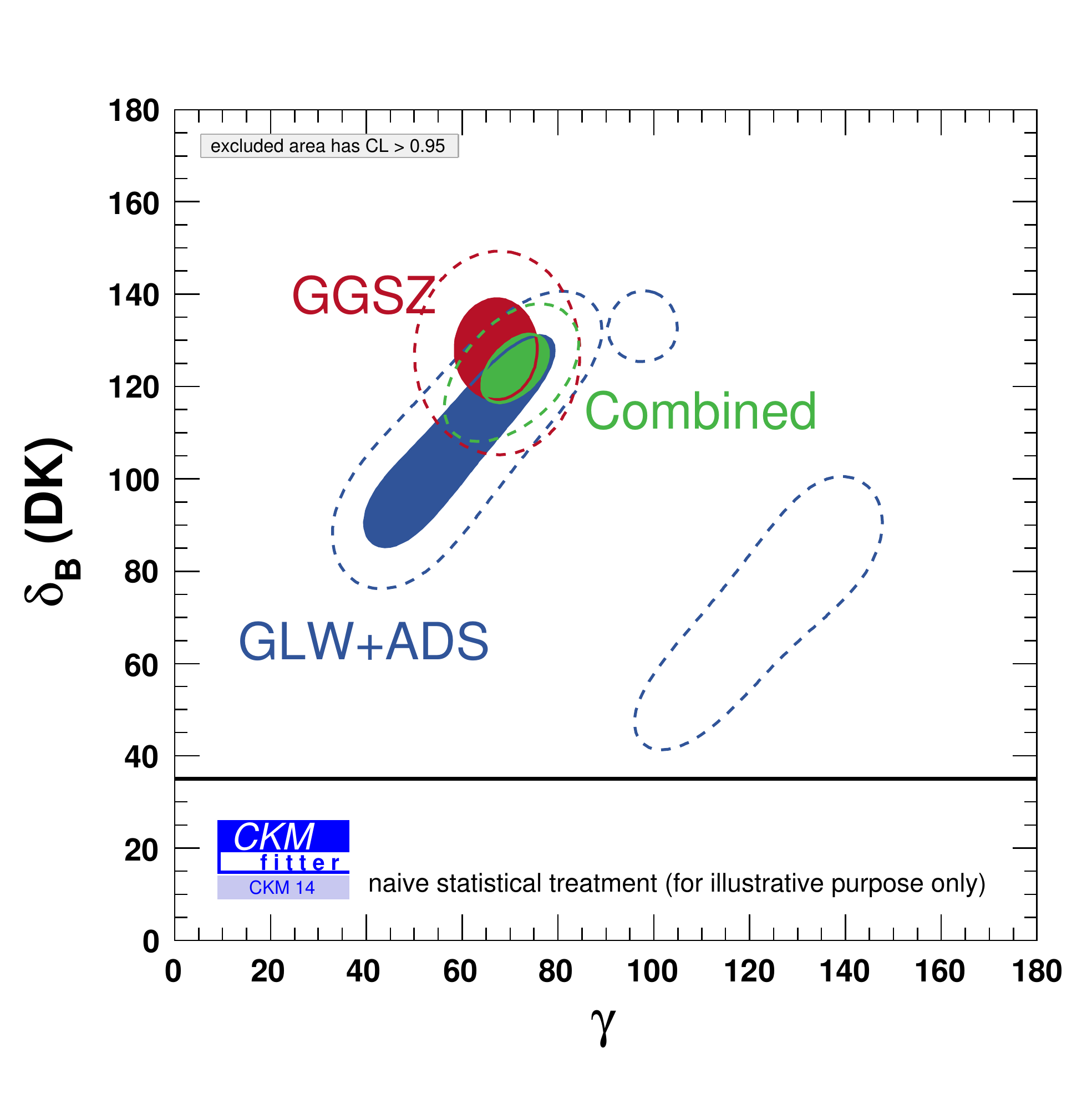}
\includegraphics[width=8cm]{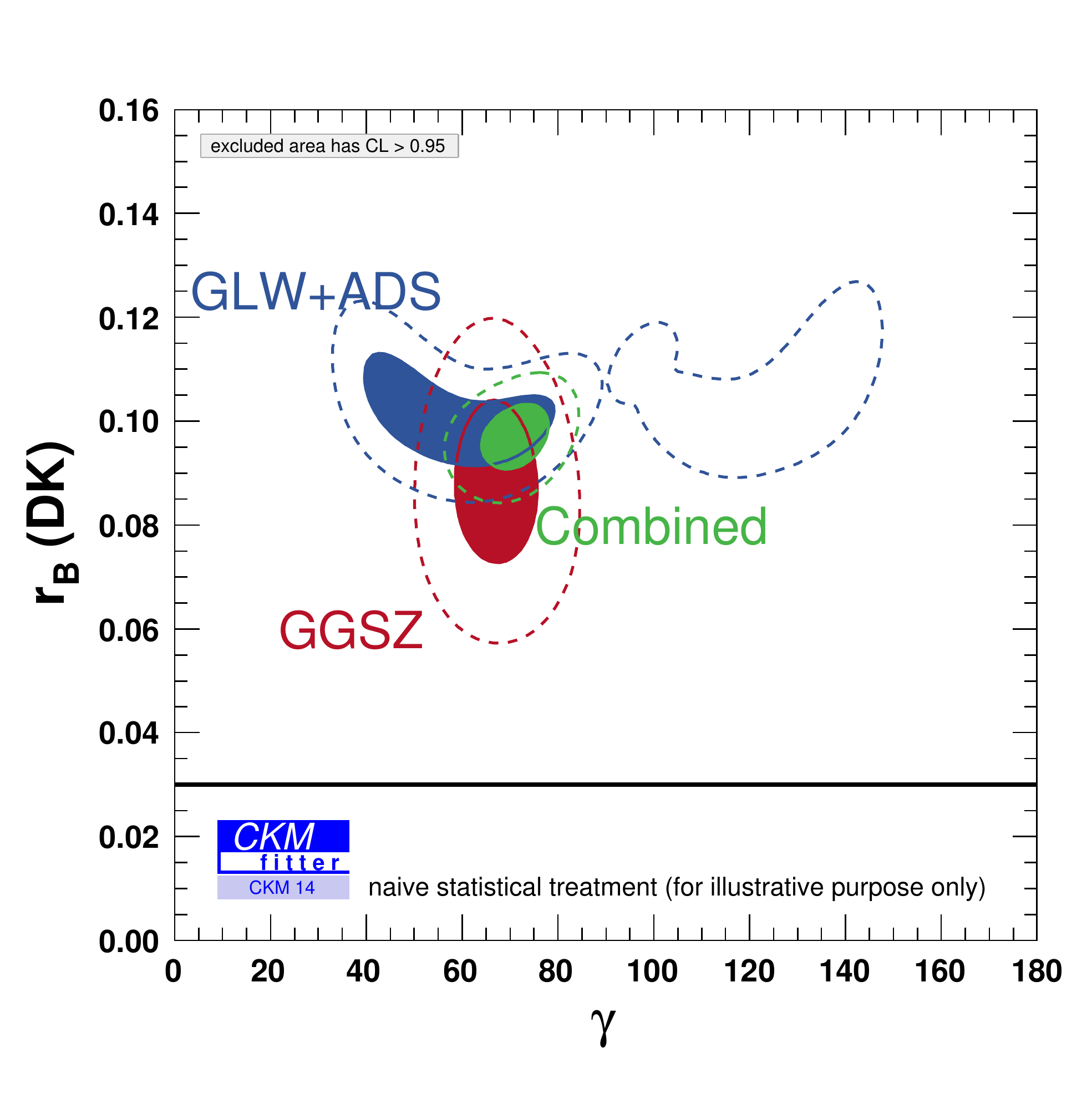} 
\caption{Inputs for $\gamma$ used in the SM global fit (top). We show the world averages for the different methods, in the $(\gamma,\delta_B)$ (middle) and $(\gamma,r_B)$ (bottom) planes.
Shaded areas (dashed lines) enclose points with $1-p < 68.3~\%$ ($95.45~\%$).
}
\label{fig:gamma}
\end{figure}

\begin{figure}
\includegraphics[width=8cm]{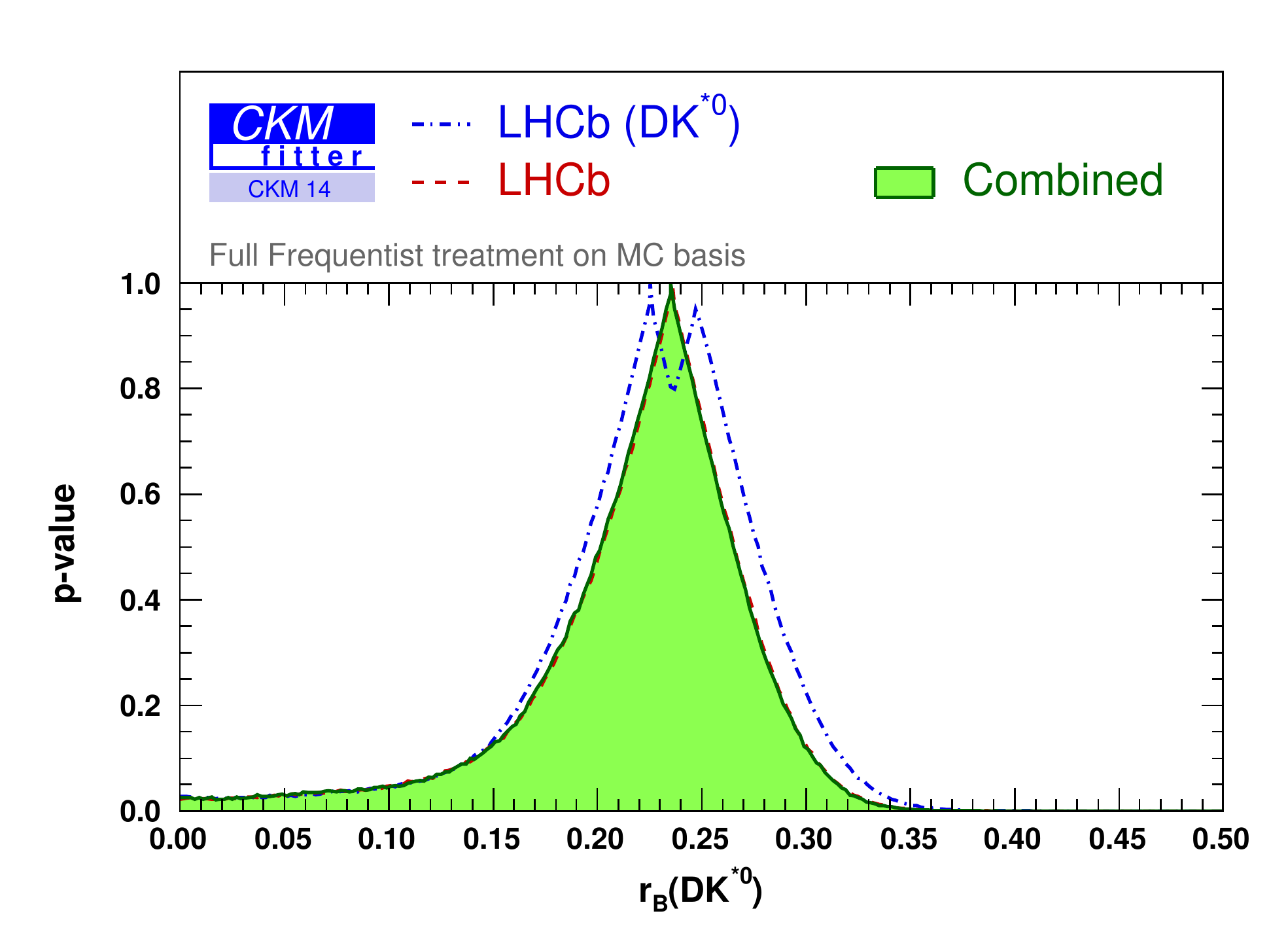} 
\caption{Constraint obtained for the $r_B$ parameter of the $B \to DK^{*0}$ mode. ``LHCb $(DK^{*0})$'' (dot-dashed line) includes only
data from $DK^{*0}$  whereas ``LHCb'' (dashed line) involves all channels (including $DK^{*0}$).}
\label{fig:gamma_dk*}
\end{figure}
For the constraint on $\gamma$, we have considered recent results from $B$-meson decays to open-charm final states, $B \to D^{(*)}K^{(*)}$. In those decays, the interference between $b \to c\bar{u}s$ and $b \to u \bar{c}s$ tree amplitudes gives access to the weak phase $\gamma$. Several methods have been proposed, which can be grouped according to the choice of the final state. 
Recent results include the updated LHCb results for the charged $B \to DK$ decay, where $D \to K_S \pi^+\pi^-$, $K_S K^+K^-$, using a 3~fb$^{-1}$ data sample~\cite{Aaij:2014uva} and for the first time, several observables, including $CP$ asymmetries, for the $B^0 \to DK^{*0}$ decays, where $D$ decays in $\pi^-K^+$, $K^-K^+$  $\pi^-\pi^+$~\cite{Aaij:2014eha}. 
Combining the experimental data for the decay modes, the world average 68.3\% $CL$ interval 
\begin{equation}
\gamma_{WA} = (73.2 {}^{+6.3}_{-7.0})^{\circ}
\end{equation}
is obtained (Fig.~\ref{fig:gamma}), as well as the hadronic parameters ($r_B$, the magnitude of the ratio of the amplitudes, and $\delta_B$, the relative strong phase between the two amplitudes) summarized in Table~\ref{tab:gamma}. Though the impact of the observables for the neutral $B$ decay $B \to DK^{*0}$ is small for the $\gamma$ measurement itself, it is worth noticing that the corresponding $r_B$ is now clearly measured away from 0, as $r_B(DK^{*0}) = 0.236 {}^{+0.043}_{-0.052}$ (Fig.~\ref{fig:gamma_dk*}). The recent measurement of LHCb with the $B_s \to D_s K$ mode~\cite{Aaij:2014fba} has not been included in our $\gamma$ average. Though very promising while using only 1 fb$^{-1}$, we estimate its impact on the $\gamma$ error to be at the order of 0.1$^\circ$.

\begin{figure*}
\vspace{-0.4cm}

\includegraphics[width=8cm]{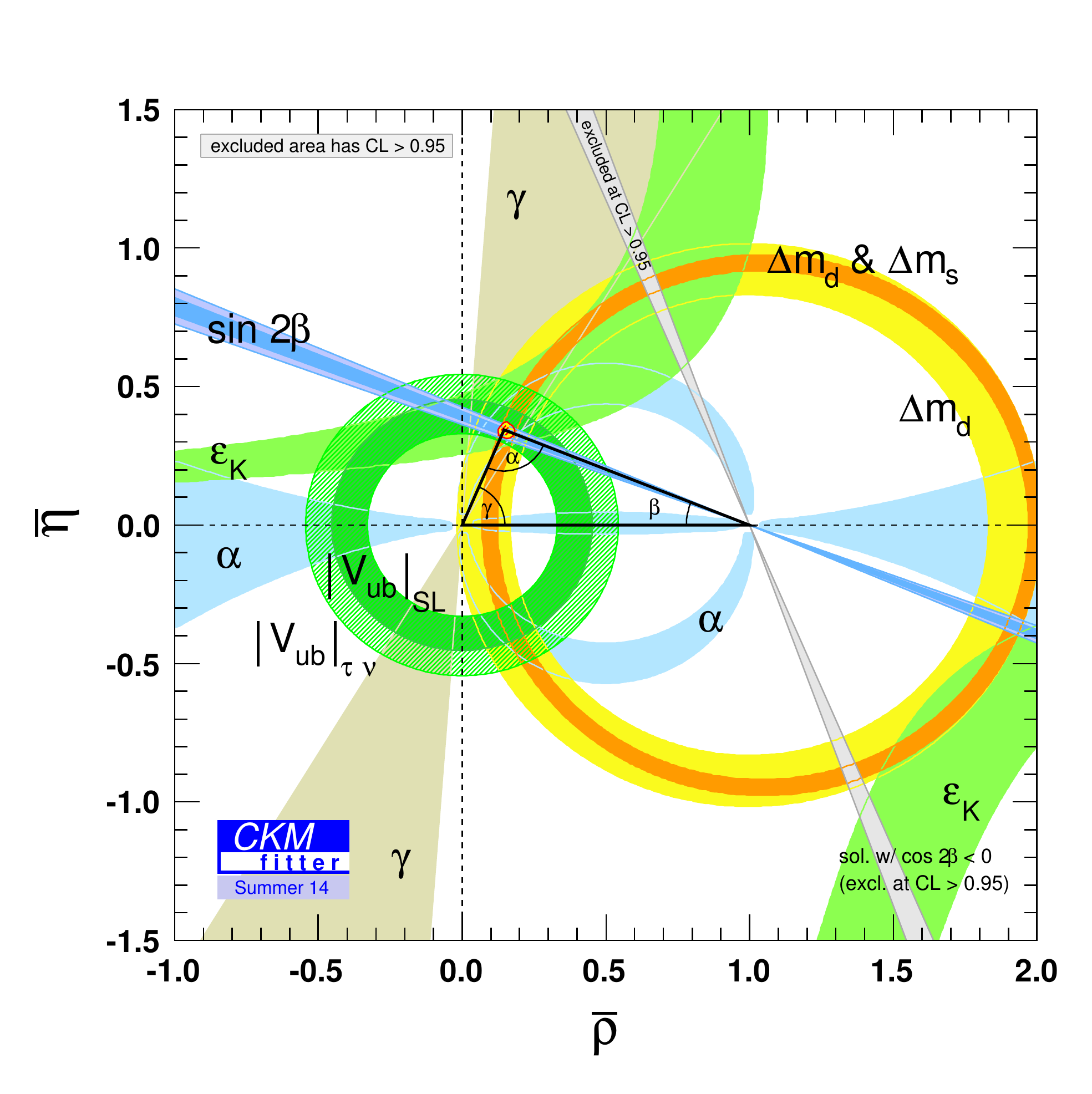}\includegraphics[width=8cm]{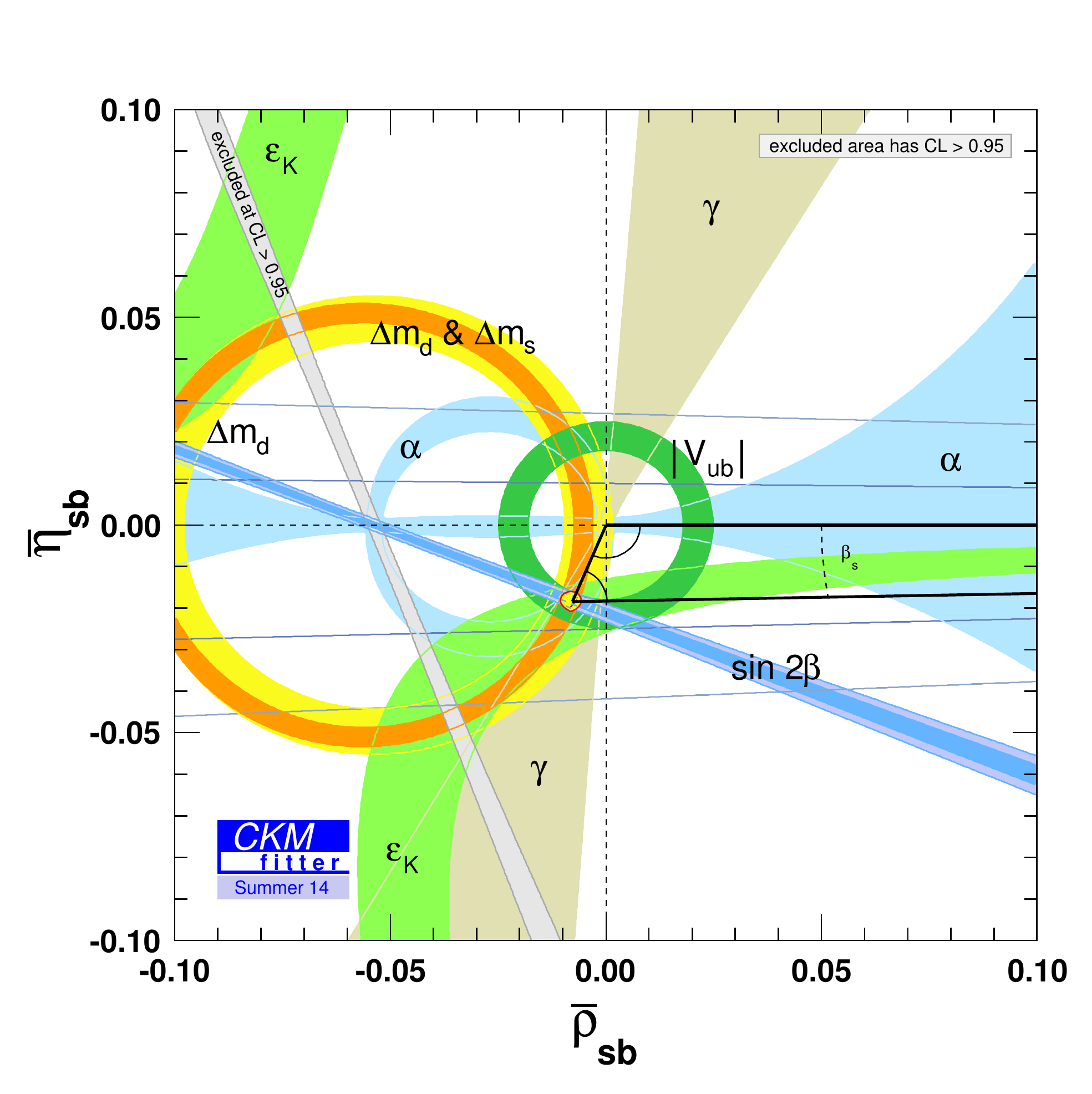}

\vspace{-0.2cm}

\includegraphics[width=8cm]{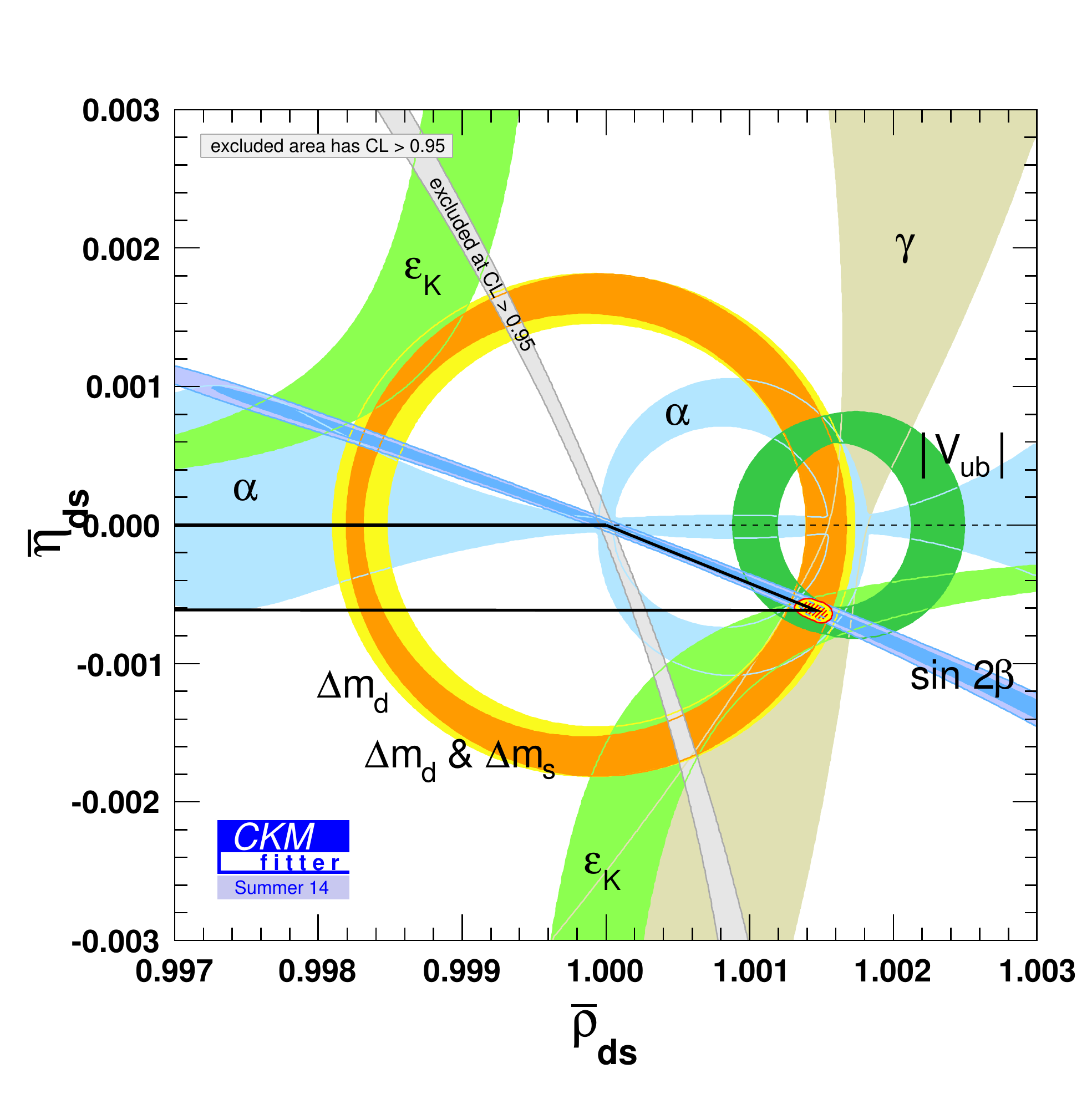}\includegraphics[width=8cm]{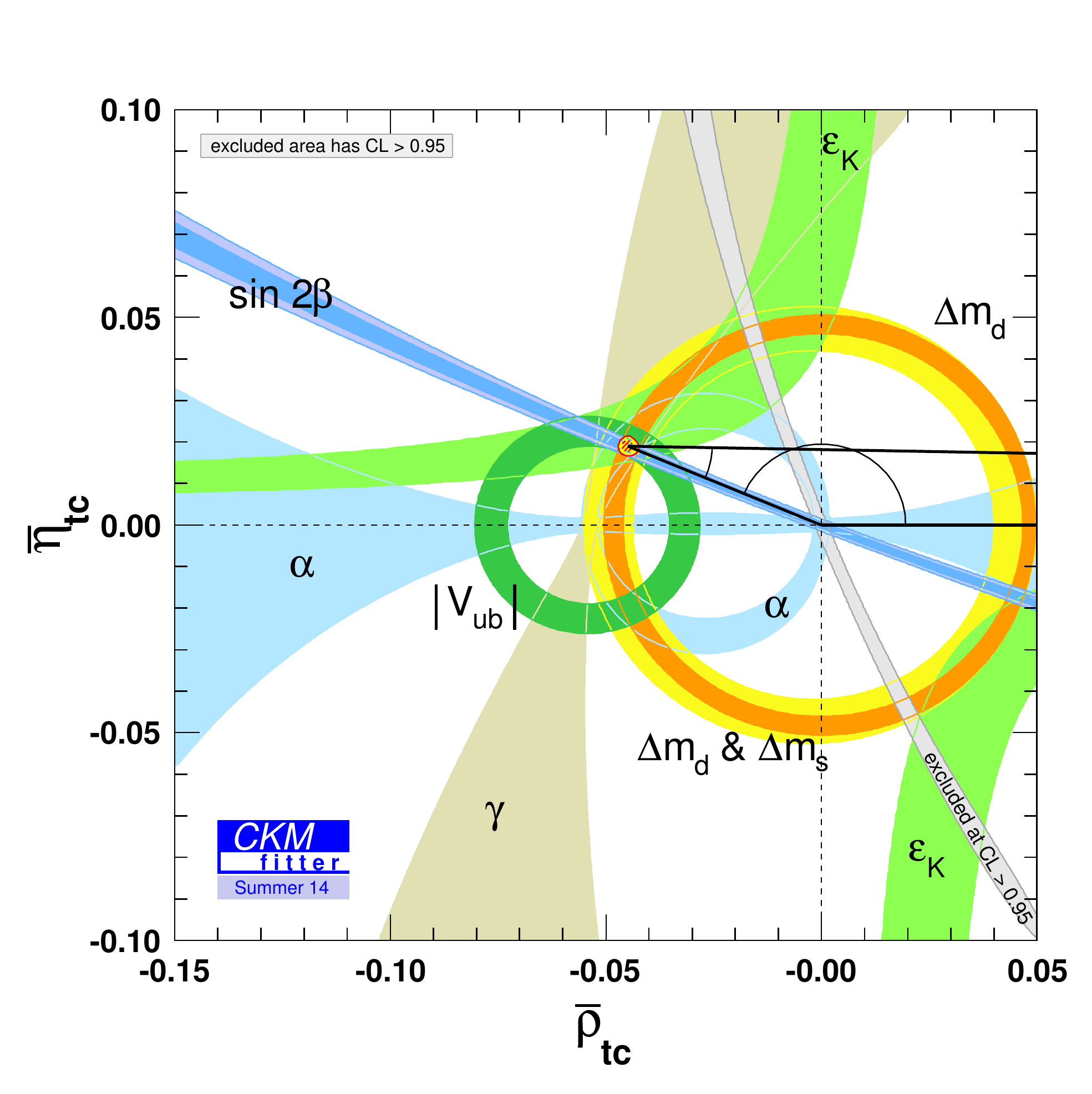}

\vspace{-0.2cm}

\includegraphics[width=8cm]{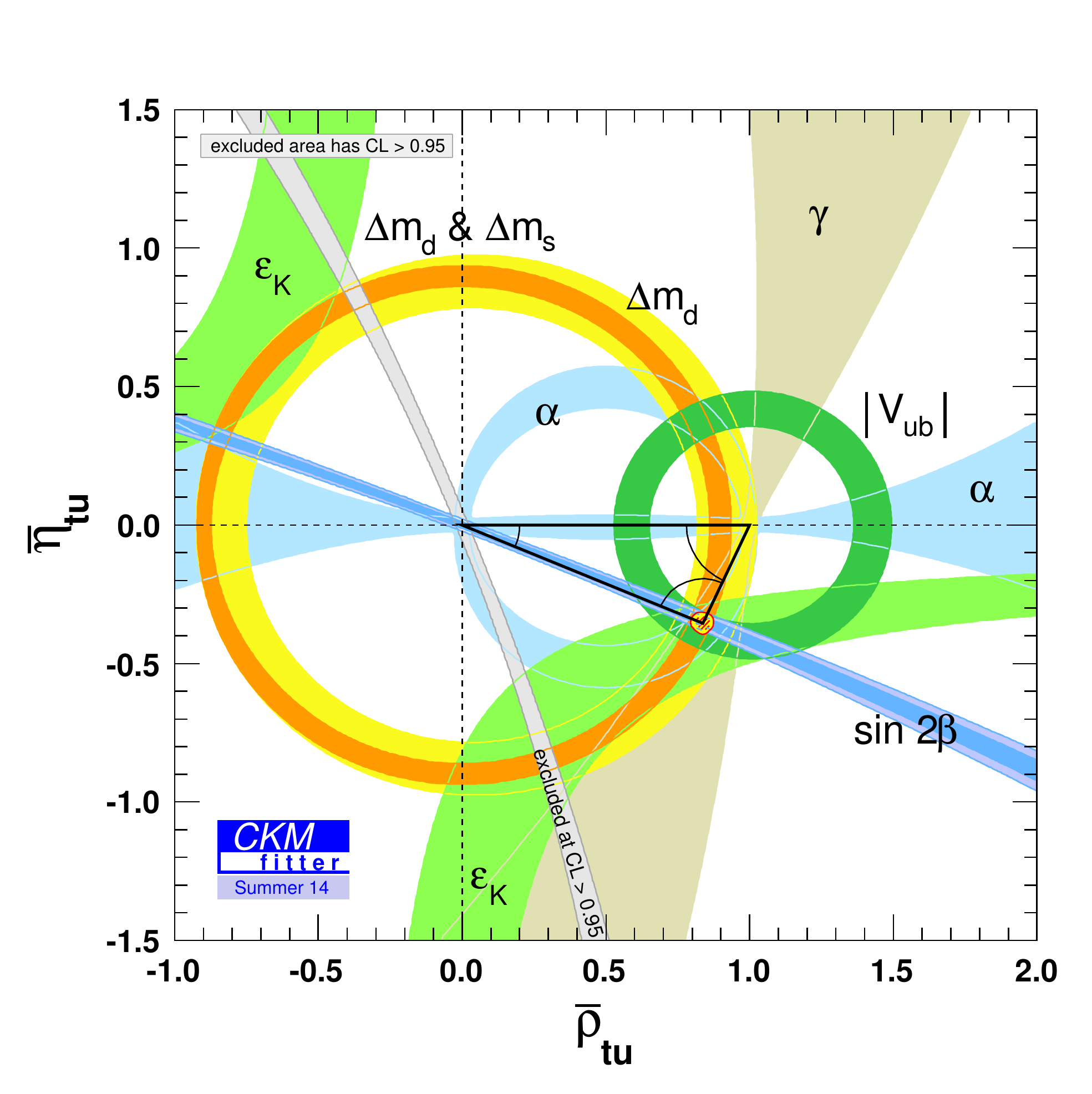}\includegraphics[width=8cm]{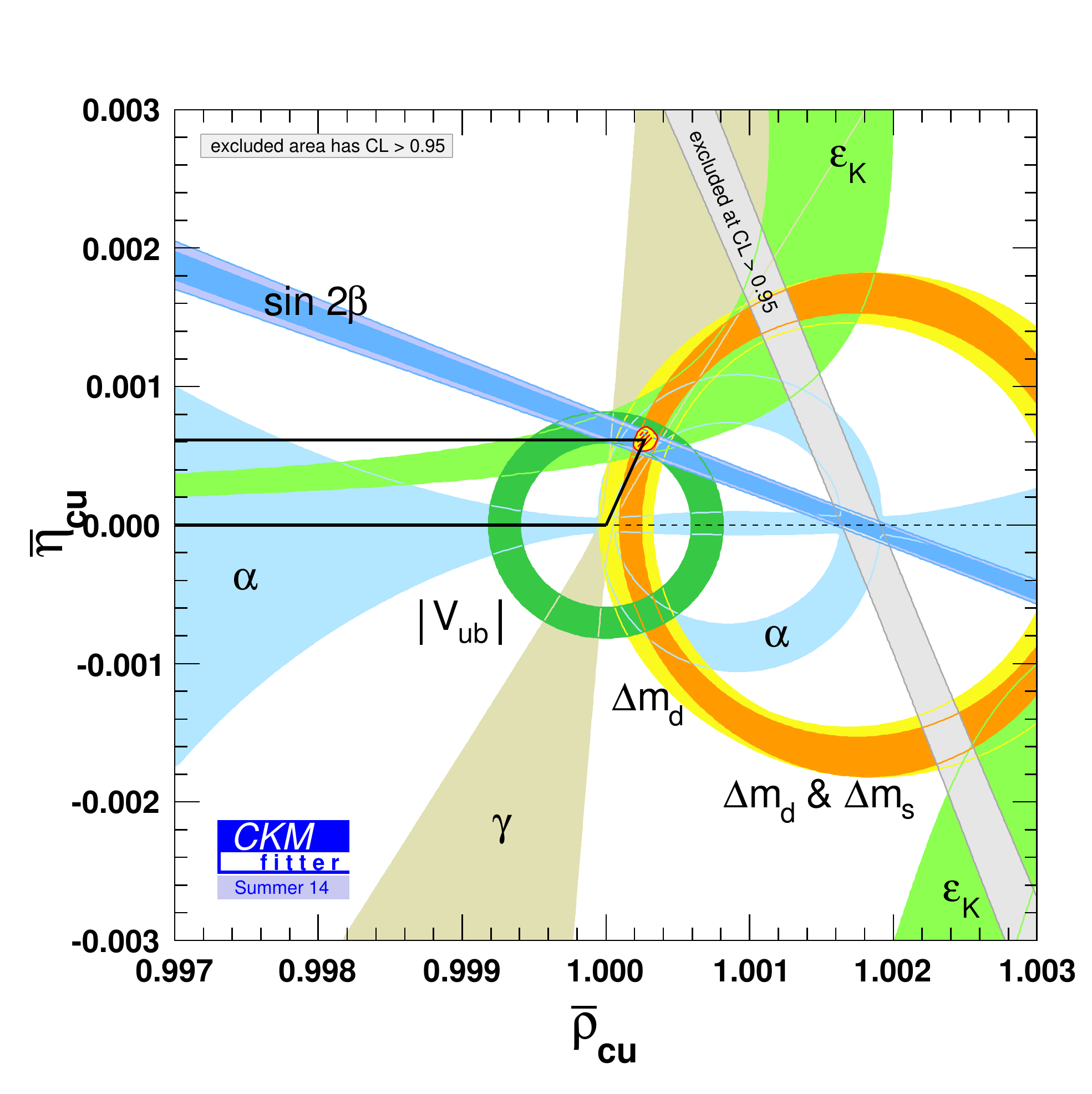}
\caption{Constraints on the CKM $(\bar\rho_{M},\bar\eta_{M})$  coordinates
with $M=db,sb,ds,ct,ut,uc$, from the global SM
CKM-fit. Regions outside the coloured areas have $1-p > 95.45~\%$. For the combined
fit the yellow area inscribed by the contour line represents points with $1-p < 95.45~\%$.
The shaded area inside this region represents points with $1-p < 68.3~\%$.\label{fig:global fit}}
\end{figure*}

\begin{table}[Htp]
\renewcommand{\arraystretch}{1.2}
\centering
\begin{tabular}{c|c}
Parameter   & Value and uncertainties  \\
\hline 
charged $B$ & \\
\hline
$r_B (DK)$      & $0.0970 {}^{+0.0062}_{-0.0063}$ \\
$\delta_B (DK)$ & $(125.4 {}^{+7.0}_{-7.8})^\circ$\\
$r_B (D^*K)$    & $0.119 {}^{+0.018}_{-0.019}$  \\
$\delta_B (D^*K)$ & $(-49 {}^{+12}_{-15})^\circ$\\
$r_B (DK^*)$   & $0.137 {}^{+0.051}_{-0.047}$ \\
$\delta_B (DK^*)$   & $(112 {}^{+32}_{-44})^\circ$ \\
\hline 
neutral $B$ & \\
\hline
$r_B (DK^*)$   & $0.236 {}^{+0.043}_{-0.052}$ \\
$\delta_B (DK^*)$   & $(336 {}^{+19}_{-23})^\circ \cup (200 {}^{+10}_{-9})^\circ$ \\
\end{tabular}
\caption{Confidence intervals for the main hadronic parameters obtained from the combination 
of the relevant BaBar, Belle and LHCb observables measured in the charged and neutral $B \to D^{(*)}K^{(*)}$ decays.}
\label{tab:gamma}
\end{table}

Other quantities which have experienced recent improvement are the branching ratios ${\cal B}(B_q\to\mu\mu)$ with $q=d,s$. ${\cal B}(B_s\to\mu\mu)$ have been observed and measured both by CMS and LHCb (at $4.3 \sigma$ and $4.0 \sigma$ respectively), leading to a rather accurate combination~\cite{Bsmumu}. There are also interesting information already available on ${\cal B}(B_d\to\mu\mu)$, even though the threshold for evidence has not been reached yet by either of the two experiments. On the theoretical side, new computations have been performed including NLO electroweak corrections and NNLO strong corrections~\cite{Bobeth:2013uxa,Hermann:2013kca,Bobeth:2013tba}, settling down issues met by earlier calculations concerning the stability with respect to higher-order corrections. In our predictions, we include the residual uncertainty of 1.5\% discussed in ref.~\cite{Bobeth:2013uxa}.
We will predict the value of the dileptonic branching ratios without time integration, which would induce a further increase of $O(\Delta \Gamma_s/\Gamma_s)$, more precisely $(1+y_s)=1.07$ discussed in refs.~\cite{DeBruyn:2012wk,DeBruyn:2012wj,DescotesGenon:2011pb}.

\section{Results of the SM global fit}

\subsection{CKM parameters and Unitarity Triangles}

The current situation of the global fit in the $(\bar\rho,\bar\eta)$ plane is indicated in Fig.~\ref{fig:global fit}. Some comments are in order before discussing the metrology of the parameters. There exists a unique preferred region defined by the entire set of observables under consideration in the global fit. This region is represented by the yellow surface inscribed by the red contour line for which the values of $\bar\rho$ and $\bar \eta$ with a $p$-value such that $1-p < 95.45~\%$. The goodness of the fit can be addressed in the simplified case where all the inputs uncertainties are taken as Gaussian, with a $p$-value found to be $66 \%$ (i.e., 0.4 $\sigma$; a more rigorous derivation of the $p$-value in the general case is beyond the scope of this article~\cite{wip}).  One obtains the following values (at 1$\sigma$) for the 4 parameters describing the CKM matrix: 
\begin{eqnarray}
A= 0.810^{\,+0.018}_{\,-0.024}\,, &\qquad&
\lambda=0.22548^{\,+0.00068}_{\,-0.00034}\,,\\
\bar\rho = 0.145^{\,+0.013}_{\,-0.007}\,, &\qquad&
\bar\eta = 0.343^{\,+0.011}_{\,-0.012}\,.
\end{eqnarray}
The various constraints can be expressed in the unitarity triangles associated with the different mesons of interest, with angles defined independently of phase conventions:
\begin{eqnarray}\label{eq:UTangles}
\alpha_{d_1d_2}&=&\mathrm{arg} \left[-\frac{V_{td_1}V_{td_2}^*}{V_{ud_1}V_{ud_2}^*} \right]~,\ \beta_{d_1d_2}=\mathrm{arg} \left[-\frac{V_{cd_1}V_{cd_2}^*}{V_{td_1}V_{td_2}^*} \right]~, \nonumber\\
\gamma_{d_1d_2}&=&\mathrm{arg} \left[-\frac{V_{ud_1}V_{ud_2}^*}{V_{cd_1}V_{cd_2}^*} \right]~, 
\end{eqnarray}
and similarly for the angles in the up sector:
\begin{eqnarray}\label{eq:UTanglesbis}
\alpha_{u_1u_2}&=&\mathrm{arg} \left[-\frac{V_{u_1b}V_{u_2b}^*}{V_{u_1d}V_{u_2d}^*} \right]~, \ \beta_{u_1u_2}=\mathrm{arg} \left[-\frac{V_{u_1s}V_{u_2s}^*}{V_{u_1b}V_{u_2b}^*} \right]~, \nonumber\\
\gamma_{u_1u_2}&=&\mathrm{arg} \left[-\frac{V_{u_1d}V_{u_2d}^*}{V_{u_1s}V_{u_2s}^*} \right]~, \end{eqnarray}
One recovers the usual $\phi_1$, $\phi_2$, $\phi_3$ and $\alpha$, $\beta$, $\gamma$ (without subscripts) for the $B_d$ Unitarity Triangle ($d_1=d,d_2=b$). In the same general way the relative coordinates of the upper appex of each triangle are defined as
\begin{eqnarray}\label{eq:UTapex}
\bar\rho_{d_1d_2}+i\bar\eta_{d_1d_2} &=& -\frac{V_{ud_1}V_{ud_2}^*}{V_{cd_1}V_{cd_2}^*}~, \nonumber\\
\bar\rho_{u_1u_2}+i\bar\eta_{u_1u_2} &=& -\frac{V_{u_1d}V_{u_2d}^*}{V_{u_1s}V_{u_2s}^*}~,
\end{eqnarray}
where again $\bar\rho+i\bar\eta\equiv\bar\rho_{db}+i\bar\eta_{db}$ refer to the $B_d$ system. 
In the $B_s$ case, $\phi_s$ can be defined as $2\beta_{sb}$. The corresponding triangles
are shown in Fig.~\ref{fig:global fit}, in particular the $(sb)$ where the constraint from $\phi_s$ is shown (but the corresponding label is not indicated).

\begin{table*}

\renewcommand\arraystretch{1.2}

\caption{Comparison between prediction and measurement of some flavour observables
in the SM. The first
  column describes the observables. The second and third columns give the measurement and the
  prediction from the global fit (not including the measurement of the quantity considered),
  respectively. The fourth column expresses the departure of the
  prediction to the measurement, when available.  
   \label{tab:pred:meas}}
  \begin{tabular}{c|cc|rl|c}
   Observable & \multicolumn{2}{c|}{Measurement} & \multicolumn{2}{c|}{Prediction}  & Pull ($\sigma$)
     \rule[-2mm]{0pt}{4ex} \\ 
  
   \hline
   \multicolumn{6}{c}{Charged Leptonic Decays \rule[-2mm]{0pt}{4ex} } \\
   \hline 
   ${\cal B}(B^+\to \tau^+\nu_{\tau}) $ & $ (10.8 \pm 2.1) \cdot 10^{-5}$  & \cite{btaunu,HFAG} 
   &  (7.58 & ${}^{+0.80}_{-0.59})\cdot 10^{-5}$   & 1.5 \\ 
   ${\cal B}(B^+\to \mu^+\nu_{\mu})$	& $< 1.0 \cdot 10^{-6}$   & \cite{HFAG} 
   &  (3.64 & ${}^{+0.27}_{-0.38})\cdot 10^{-7}$   & - \\ 
   ${\cal B}(D_s^+\to \tau^+\nu_{\tau}) $	& $(5.55 \pm 0.24)\cdot 10^{-2}$ &  \cite{HFAG} 
   &  (5.19  & ${}^{+0.02}_{-0.12})\cdot 10^{-2}$   & 1.5 \\ 
   ${\cal B}(D_s^+\to \mu^+\nu_{\mu}) $	& $(5.57  \pm 0.24)\cdot 10^{-3}$ &    \cite{HFAG} 
   &  (5.31  & ${}^{+0.02}_{-0.09})\cdot 10^{-3}$   & 1.1 \\ 
   ${\cal B}(D^+\to \mu^+\nu_{\mu}) $ & $(3.74  \pm 0.17)\cdot 10^{-4}$ &   \cite{HFAG}
   & (3.91  & $\pm 0.11)\cdot 10^{-4}$   & 0.6 \\ 
   \hline
   \multicolumn{6}{c}{Neutral Leptonic $B$ decays  \rule[-2mm]{0pt}{4ex}} \\
   \hline
  ${\cal B}(B^0_s \to \tau^+\tau^-)$	&  -  &
   &  (6.92  & ${}^{+0.41}_{-0.39})\cdot 10^{-7}$   &  -\\ %
   ${\cal B}(B^0_s \to \mu^+\mu^-)$	& $(2.8^{+0.7}_{-0.6})\cdot 10^{-9}$ & \cite{Bsmumu}
   &  (3.34  & ${}^{+0.13}_{-0.25})\cdot 10^{-9}$   &  1.0 \\ %
  ${\cal B}(B^0_s \to e^+ e^-)$	&  $< 2.8\cdot 10^{-7}$    &  \cite{HFAG} 
   &  (7.64  & ${}^{+0.46}_{-0.43})\cdot 10^{-14}$   & - \\%
   ${\cal B}(B^0_d \to \tau^+\tau^-)$	& $<4.1  \cdot 10^{-3}$  & \cite{HFAG} 
   &  (2.05 & ${}^{\,+0.13}_{\,-0.14})\cdot 10^{-8}$   & - \\ %
  ${\cal B}(B^0_d \to \mu^+\mu^-)$	& $(3.6 {}^{+1.6}_{-1.4})\cdot 10^{-10}$ & \cite{Bsmumu} 
   &  (0.98  & ${}^{+0.06}_{-0.07})\cdot 10^{-10}$   & -\\ %
   ${\cal B}(B^0_d \to e^+ e^-)$	& $<$ $8.3\cdot 10^{-9}$  &  \cite{HFAG} 
   &  (2.29  & ${}^{+0.14}_{-0.16})\cdot 10^{-15}$   &  -\\ %

   \hline
   \multicolumn{6}{c}{$\bbq$ mixing observables \rule[-2mm]{0pt}{4ex}} \\
   \hline
     $\Delta\Gamma_s$  (ps$^{-1}$) & $0.081\pm 0.006$  & \cite{HFAG} 
   & 0.120 & ${}^{+0.043}_{-0.045}$   & 0.1 \\
  $a_{\rm SL}^d  $	& $(1 \pm 20)\cdot 10^{-4}$  & \cite{HFAG} 
     & (\,$-6.5$ & ${}^{+1.8}_{-1.9}\ \,)\cdot 10^{-4}$   & 0.3 \\ 
   $a_{\rm SL}^s $	& $(-48\pm 48)\cdot 10^{-4}$  &\cite{HFAG}
   & (0.29  & ${}^{+0.08}_{-0.08})\cdot 10^{-4}$   & 1.0 \\ 
      $A_{\rm SL} $	& $(-47\pm 17)\cdot 10^{-4}$  &\cite{Abazov:2013uma}
   & ($-3.4$  & ${}^{+1.0}_{-1.1})\cdot 10^{-4}$   & 2.7 \\ 
 $\sin(2\beta)$ & 0.682 $\pm$ 0.019 &   \cite{HFAG} 
   &  0.771 & ${}^{+0.017}_{-0.041}$   & 1.7 \\
$\phi_s$ & $-0.015\pm 0.035$  &\cite{HFAG} 
   & $-0.0365$ & ${}^{+0.0013}_{-0.0012}$   & 0.6 \\
     \hline
   \multicolumn{6}{c}{Rare $K$ decays \rule[-2mm]{0pt}{4ex}} \\
   \hline
  ${\cal B}(K^+ \to \pi^+\nu\bar\nu)$	& $(1.75^{+1.15}_{-1.05})\cdot 10^{-10}$   & \cite{E949}
   &  (0.85  & ${}^{+0.13}_{-0.12})\cdot 10^{-10}$   & 0.7 \\
  ${\cal B}(K_L \to \pi^0\nu\bar\nu) $	&  - & 
   &  (0.28  & ${}^{+0.04}_{-0.05})\cdot 10^{-10}$  & - \\
 \end{tabular}
\end{table*}

\subsection{Comments and predictions}

As underlined above, the overall consistency seen among the constraints allows us to perform the metrology of the CKM parameters and to give predictions for any CKM-related observable within the SM.  Let us add that the existence of a  $1-p < 95.45~\%$ region in the $({\bar\rho,\bar \eta})$ plane is not equivalent to the statement that each individual constraint lies in the global range of $1-p< 95.45~\%$.  Each comparison between the prediction issued from the fit and the corresponding measurement constitutes a null test of the SM hypothesis. 

\begin{figure}[t]
\hbox{\hspace{-0.65cm}\includegraphics[width=10cm]{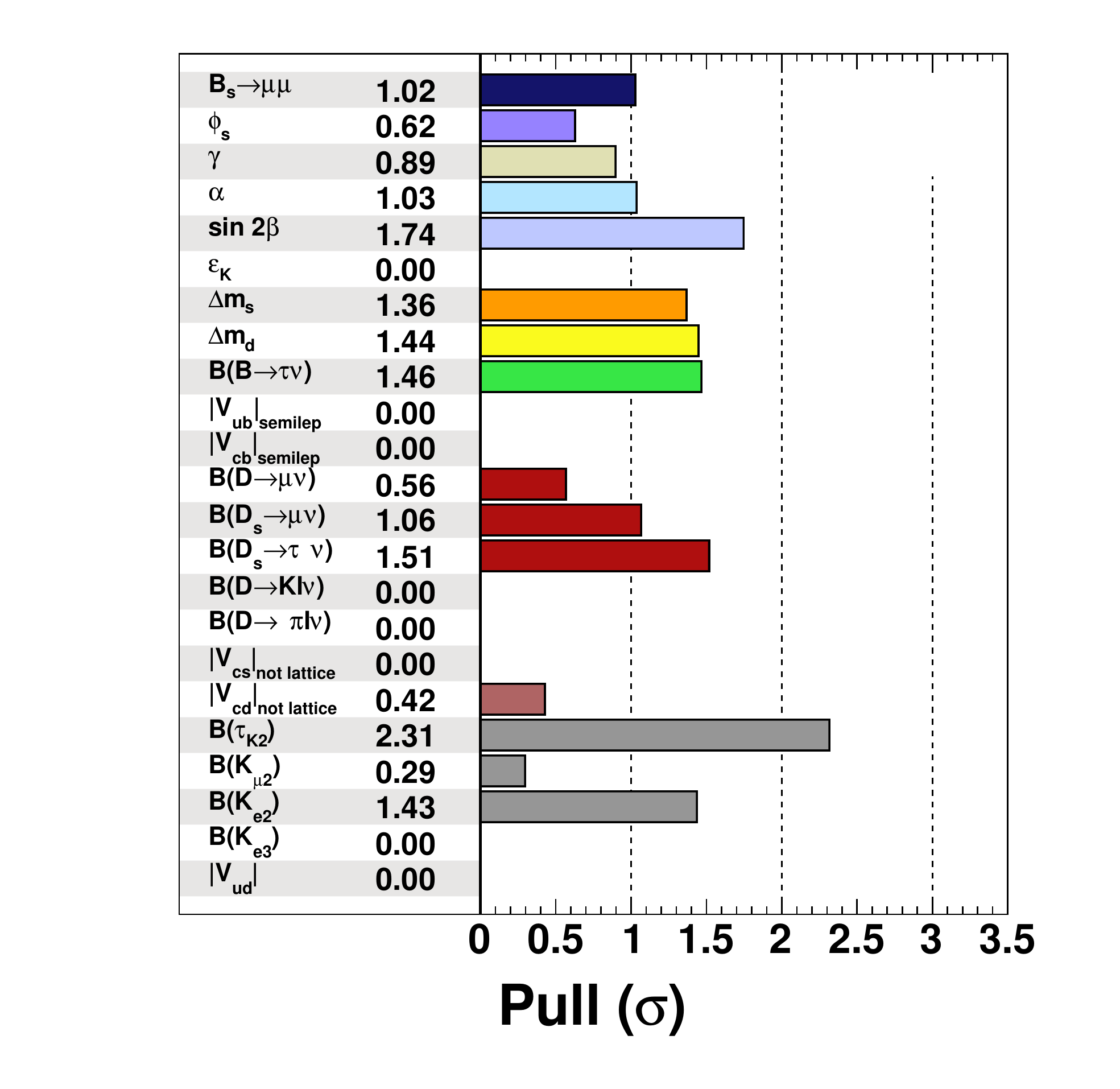}}
\caption{Pulls for the SM global fit obtained by comparing the value of $\chi^2_{\rm min}$ with and without including the measurement of the quantity. Notice that the different pulls are not necessarily independent.\label{fig:Pulls}}
\end{figure}

Some of the corresponding pulls are reported in Table~\ref{tab:pred:meas} and shown in Fig.~\ref{fig:Pulls}, showing that there is no sign of discrepancy with our set of inputs. 
One should also notice that some of the quantities included in our fit have only a limited impact on the outcome. This is for instance the case for quantities where the measurement is compatible, but less precise than the SM prediction, like $\phi_s$, ${\cal B}(B_s\to\mu\mu)$, or semileptonic and leptonic $D_{(s)}$ decays.
In Table~\ref{tab:pred:meas}, we also include observables that were not used as input constraints, either because they are not measured at a sufficient accuracy yet, e.g., ${\cal B}(B_d\to\ell^+\ell^-)$, or because the control on the theoretical uncertainties remains under discussion, e.g., $\Delta\Gamma_s$. The corresponding predictions can then be directly compared with their experimental measurements (when they are available).

\begin{figure}
\includegraphics[width=8cm]{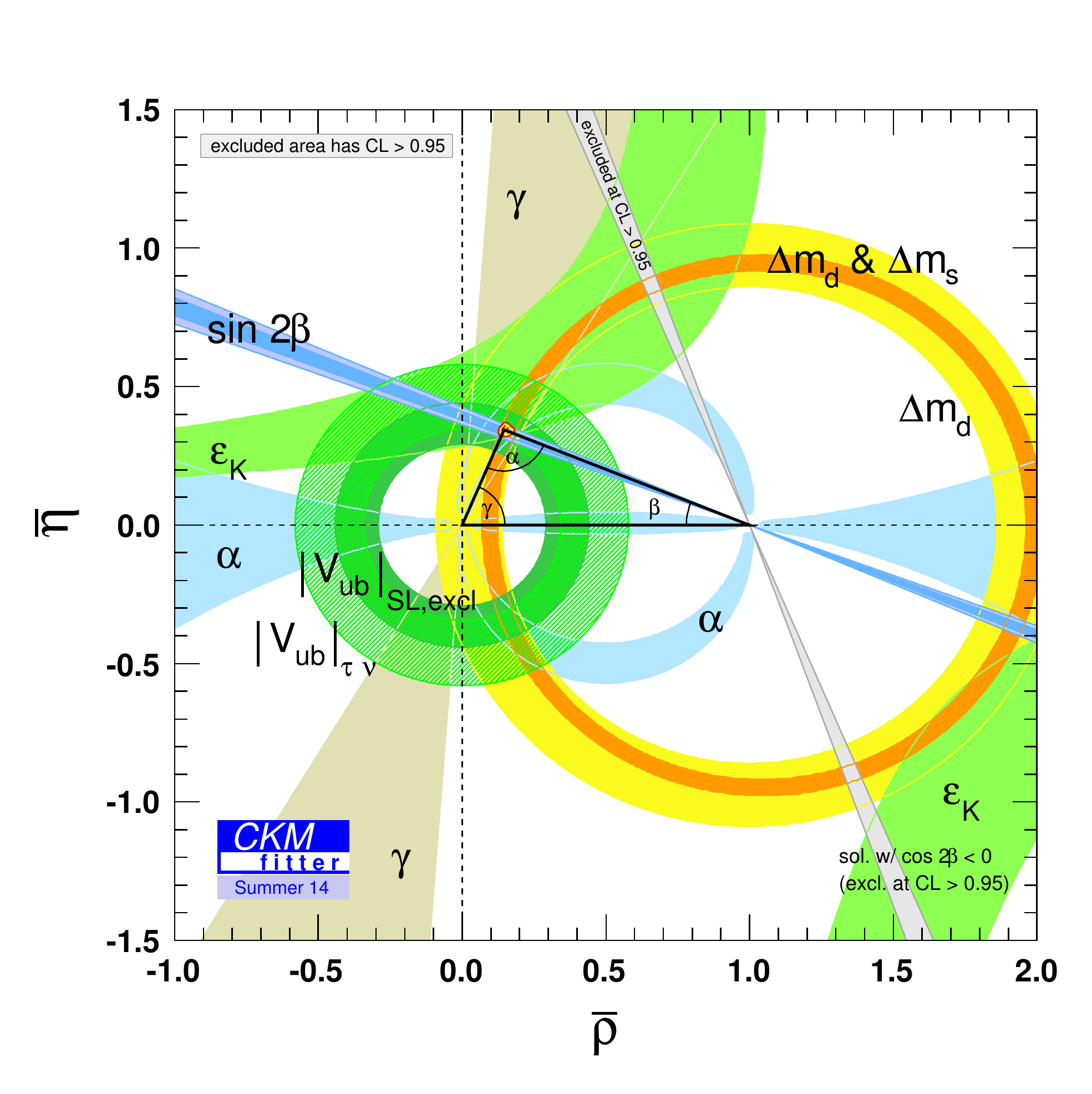}
\includegraphics[width=8cm]{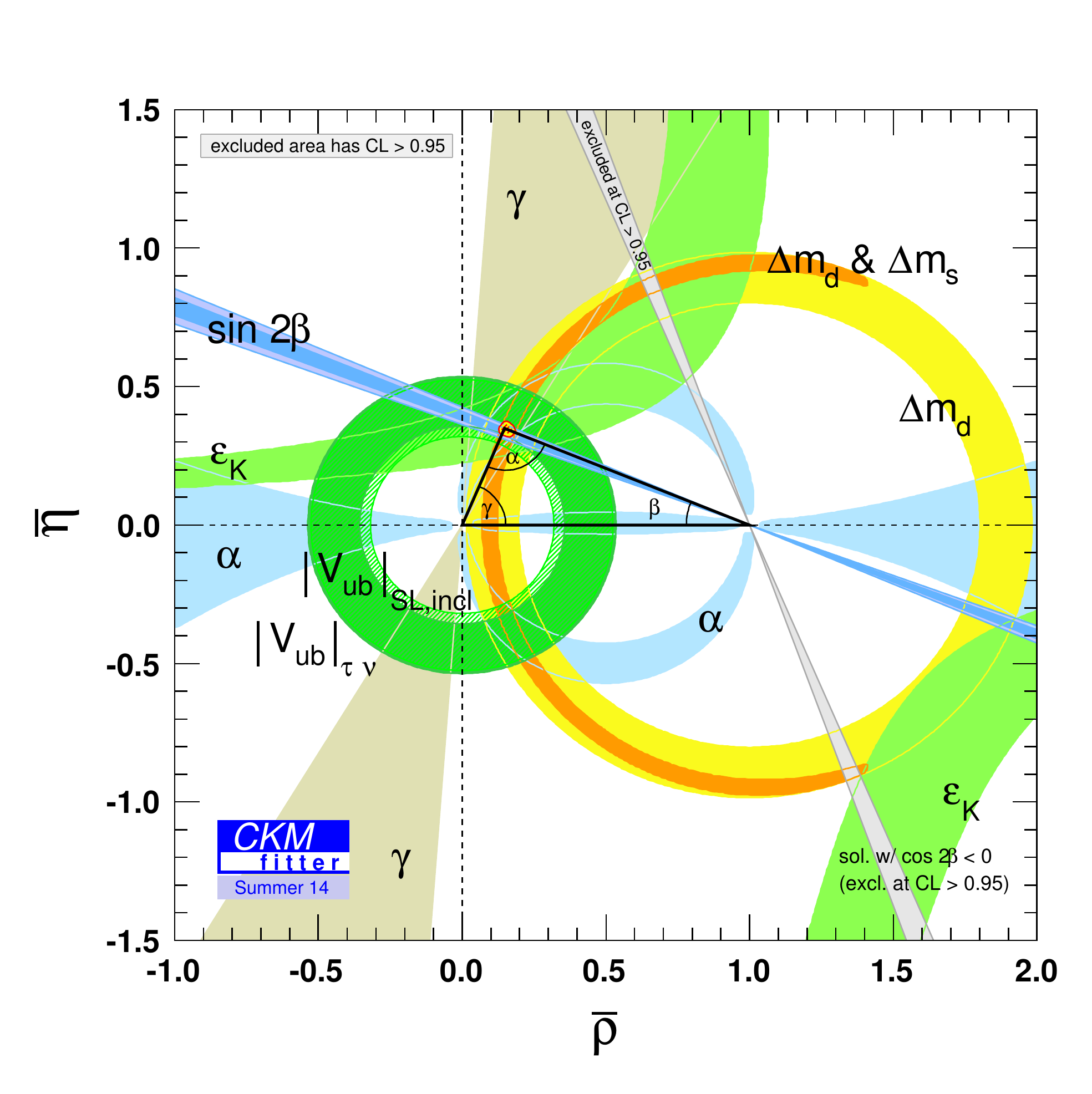}
\caption{Constraints on the CKM $(\bar\rho,\bar\eta)$ coordinates from the global SM CKM-fit using only exclusive (top) and inclusive (bottom) determinations of $|V_{ub}|$ and $|V_{cb}|$ from semileptonic decays as inputs.}
\label{fig:global_fit_inclexcl}
\end{figure}

Before moving to specific observables and correlations, we briefly discuss the lasting discrepancy between determinations of $|V_{ub}|$ and $|V_{cb}|$ using exclusive and inclusive semileptonic decays. As indicated previously, the global SM fit is based on an average for the two matrix elements, taking into account the differences between statistic and systematic errors. In Fig.~\ref{fig:global_fit_inclexcl}, we illustrate the results obtained by considering only exclusive (top) or inclusive (bottom) determinations for both $|V_{ub}|$ and $|V_{cb}|$. As expected, the constraint from $|\epsilon_K|$ changes significantly due to the variation in $|V_{cb}|$, whereas the $|V_{ub}|$ constraint from $B\to\tau\nu$ is found in better agreement with the inclusive  input than the exclusive one. An additional interesting feature in the inclusive case is the appearance of a partial
ring from the combined contribution of $\Delta m_d$ and $\Delta m_s$. This feature appeared already in the SM fit for Summer 2012, and can be explained by the fact that this constraint combines constraints on $|V_{td}|^2=A^2\lambda^6[(1-\bar\rho)^2+\bar\eta^2+O(\lambda^4)]$ (yielding a ring in the $(\bar\rho,\bar\eta)$ plane) and on $|V_{ts}|^2=A^2\lambda^4[1-\lambda^2(1-2\bar\rho)^2+O(\lambda^4)]$ (cutting too large values of $\bar\rho$).
The overall agreement between the various constraints remains excellent in both inclusive and exclusive fits, with very little variation in the global $p$-value at the best-fit point and the confidence interval for the four Wolfestein parameters with respect to the global SM fit obtained from an average of inclusive and exclusive values for $|V_{ub}|$ and $|V_{cb}|$.

\subsection{Specific observables and correlations}

\begin{figure}
\includegraphics[width=8cm]{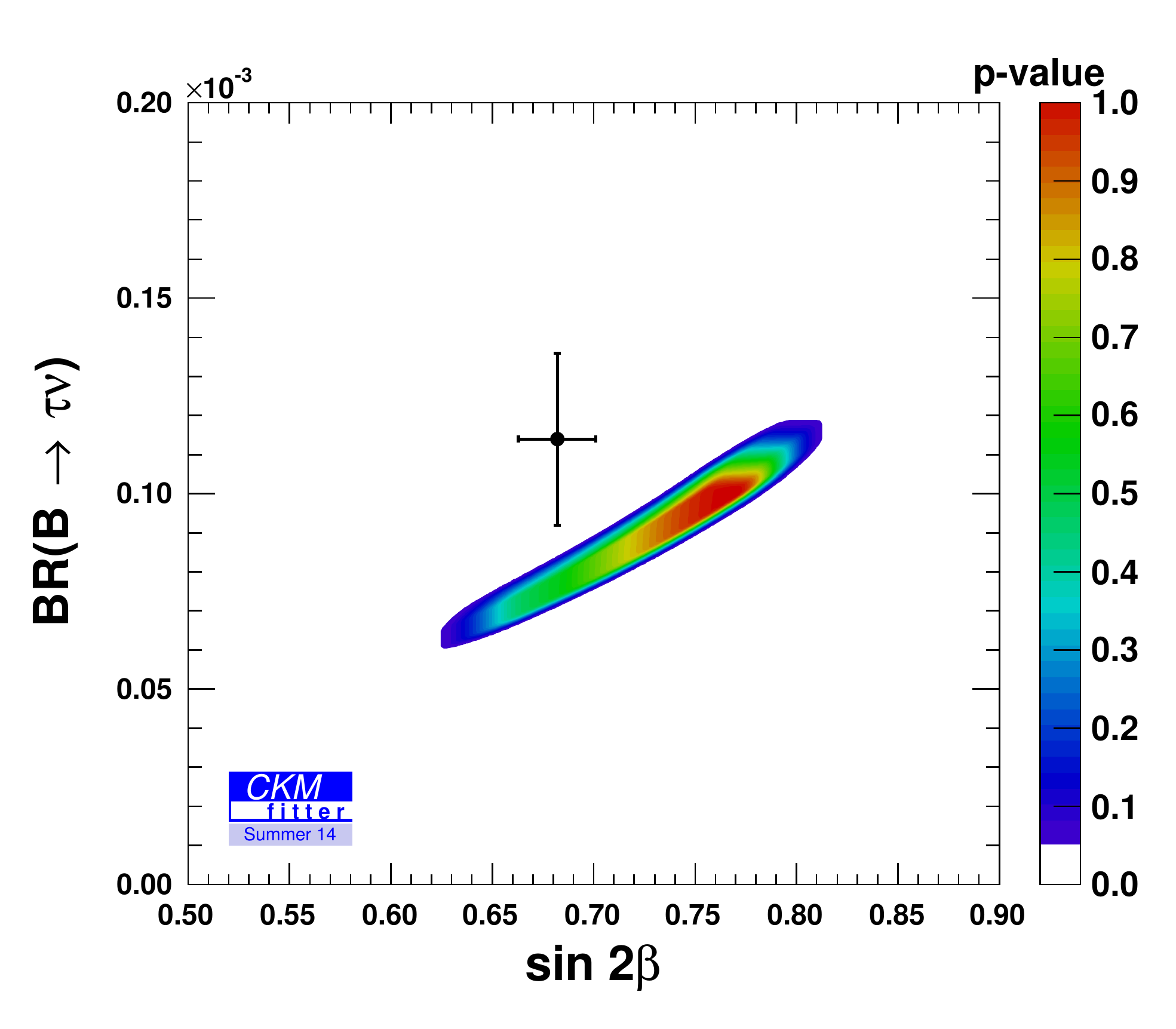}
\caption{Prediction on ${\cal B}(B\to\tau \nu)$ and $\sin 2\beta$ coming from the global fit (without the corresponding inputs) compared to current experimental information (cross). Regions outside the coloured areas are excluded at $1-p > 95.45~\%$. \label{fig:sin2bBtaunu}}
\end{figure}

We focus now on some specific observables and their correlations. A first example is given by
the two-dimensional comparison for ${\cal B}(B\to \tau\nu)$ and $\sin 2\beta$ in Fig.~\ref{fig:sin2bBtaunu}, showing that the discrepancy that used to affect
the SM global fit~\cite{Lenz:2010gu} has now been alleviated to a large extent (remaining only at $1.6 \sigma$). 
As discussed in ref.~\cite{Lenz:2010gu}, this discrepancy had an experimental origin, and it has been reduced thanks to the addition of new data (the remaining discrepancy is driven by the larger BaBar result compared to Belle measurement).

\begin{figure}
\includegraphics[width=8cm]{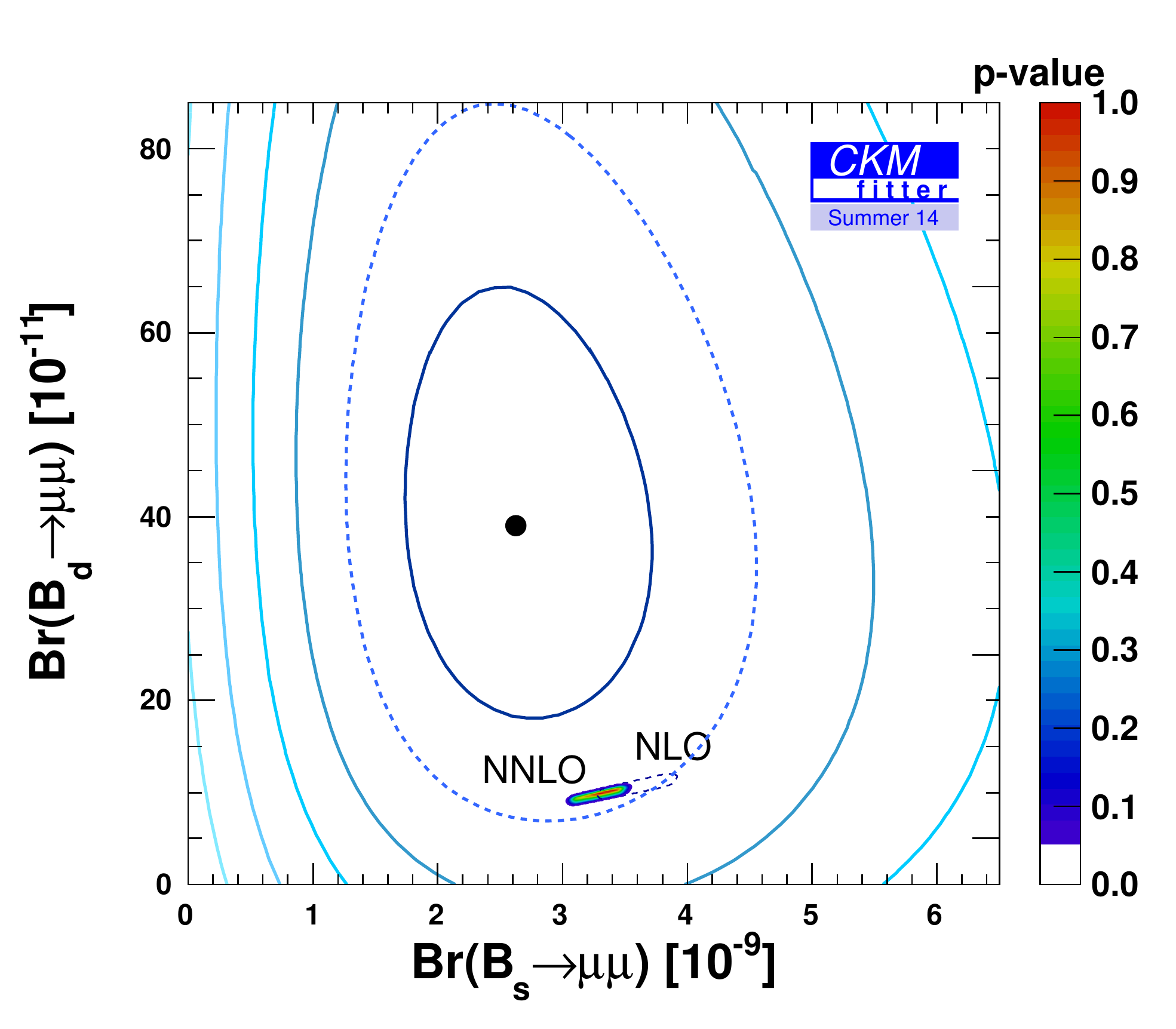}
\caption{Prediction on the two dileptonic branching ratios ${\cal B}(B_s\to\mu\mu)$ and ${\cal B}(B_d\to\mu\mu)$ coming from the global fit (without input on dileptonic branching ratios) compared to current experimental information~\cite{Bsmumu}. ${\cal B}(B_s\to\mu\mu)$ is shown removing the $(1+y_s)=1.07$ increase due to time integration. The NNLO computation in ref.~\cite{Bobeth:2013uxa} is indicated in colours, whereas the NLO computation used in ref.~\cite{Buchalla:1995vs} is  the region delimited by the dashed line. Regions outside the coloured areas are excluded at $1-p > 95.45~\%$. The oval contours are the experimental $1,2,3\ldots \sigma$ confidence regions~\cite{Bsmumu}.   \label{fig:BdBstomumu}}
\end{figure}

One can also consider ${\cal B}(B_{d,s}\to\mu\mu)$ as shown in Fig.~\ref{fig:BdBstomumu}, showing the confidence contours from the combination of CMS and LHCb~\cite{Bsmumu}. 
One notices that NLO and NNLO predictions follow the same correlation: the ratio of branching ratios is driven by $f_{B_s}/f_{B_d} |V_{ts}/V_{td}|$ which is determined to a high accuracy in the global fit.  On the other hand, the NNLO prediction is both lower and more accurate than the NLO case, in agreement with the results in ref.~\cite{Bobeth:2013uxa}. This highlights the importance of a precise measurement of this observable, e.g., at Belle-II.

\begin{figure}
\includegraphics[width=8cm]{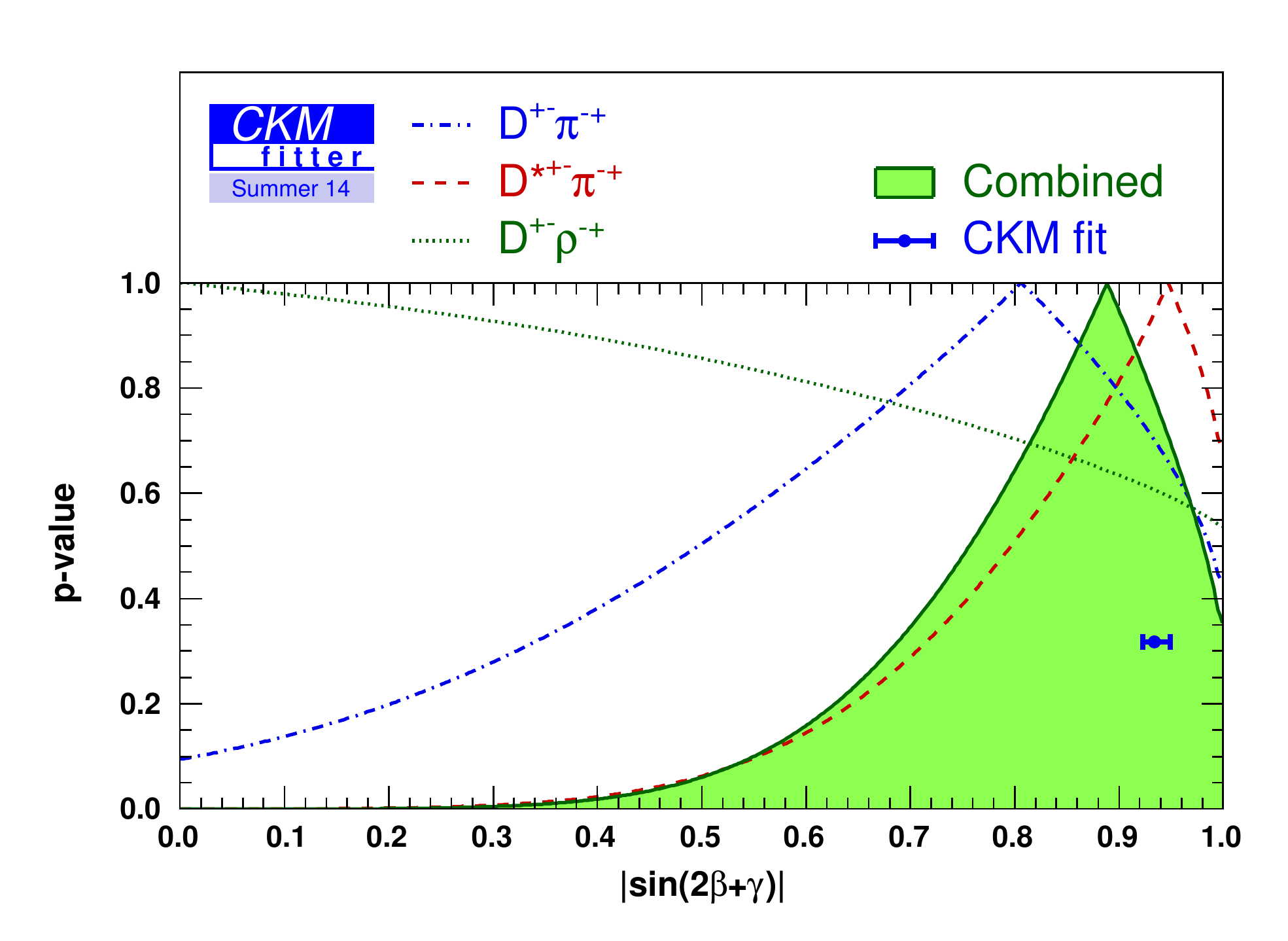}
\includegraphics[width=8cm]{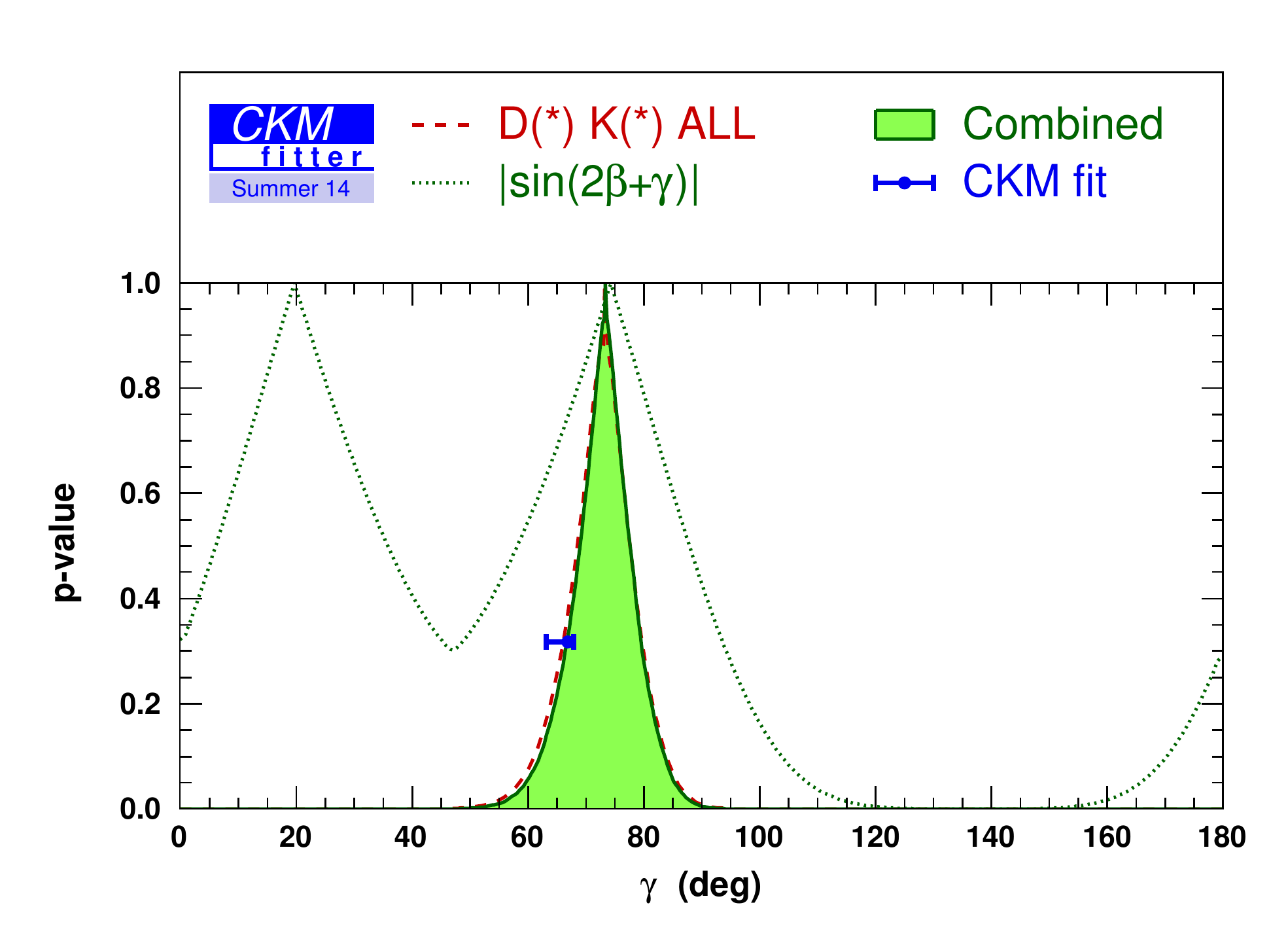}
\caption{Combined constraint on $2\beta+\gamma$ using relevant observables measured in the $B \to D\pi, D^{*}\pi$ (top) and $D\rho$ (bottom) decays.}
\label{fig:sin2bpg}
\end{figure}

The study of the time-dependent decay rates of $B \to D^\pm\pi^\mp$, $D^{*\pm}\pi^\mp$ and $D^\pm\rho^\mp$ provides a measure of $r \sin(2\beta+\gamma)$, where $r$ is the ratio of the magnitudes of the doubly-Cabibbo-suppressed and Cabibbo-favoured amplitudes \cite{Dunietz:1997in}. Because of the smallness of this ratio for the three modes, one has to extract them from $B^0 \to D_s^{(*)+}h^-$ decays assuming $SU(3)$ flavour symmetry (allowing for $SU(3)$ breaking at the level of  $1\pm 0.10\pm 0.05$). Another additional input needed is the ratio of decay constants for excited mesons: $f_{D_s}^*/f_{D}^*=1.16\pm 0.02\pm 0.06$~\cite{Becirevic:2012ti}.
Combining those observables, we obtain a constraint on$|\sin(2\beta+\gamma)|$, which corresponds to a lower limit $|\sin (2\beta+\gamma)| > 0.69$ at 68\% $CL$ (Fig.~\ref{fig:sin2bpg}). 

\section{New Physics in $\Delta F=2$}

\subsection{Additional inputs and parameters}

\begin{table}[t]
\renewcommand{\arraystretch}{1.2}
\centering
\begin{tabular}{c|c|c}
Observable                                      & Value and uncertainties                      & Ref.\              \\
\hline
$A_\text{SL}$                                   & $(-47 \pm 17)\times 10^{-4}$  & \cite{Abazov:2013uma} \\
$a_{\rm SL}^s $ &  $(1\pm 20)\cdot 10^{-4}$ & \cite{HFAG}\\
$a_{\rm SL}^d $ & $(-48\pm 48)\cdot 10^{-4}$ &\cite{HFAG}\\
$\Delta\Gamma_s$ & $0.081\pm 0.008$ & \cite{HFAG}\\
$\tilde{B}_{S,B_s}/\tilde{B}_{S,B_d}$ & $1.01\pm 0.02\pm 0.02$ & \cite{Carrasco:2013zta}\\
$\tilde{B}_{S,B_s}(m_b)$ & $0.89\pm 0.10\pm 0.09$ & \cite{Carrasco:2013zta}\\  
\end{tabular}
\caption
{Experimental and theoretical inputs inputs modified 
compared to Ref.~\cite{Lenz:2010gu,Lenz:2012az} and used in our fits for NP in $\Delta F=2$.}
\label{tab:InputsNP}
\end{table}

As discussed in refs.~\cite{Soares:1992xi,Goto:1995hj,Silva:1996ih,Grossman:1997dd,Bona:2005eu,Ligeti:2006pm,Bona:2007vi,Lenz:2010gu,Lenz:2012az,Charles:2013aka}, neutral-meson mixing is a particularly interesting probe of NP. The evolution of the $B_q\bar{B}_q$ system is described through a quantum-mechanical hamiltonian $H=M^q-i\Gamma^q/2$ as the sum of two hermitian ``mass'' and ``decay'' matrices. 
\bbq $(q=d,s)$ oscillations involve the off-diagonal elements 
$M_{12}^q$ and $\Gamma_{12}^q$, respectively.
One can fix the three physical quantities $|M_{12}^q|$,
  $|\Gamma_{12}^q|$ and $\phi_q=\arg(-M_{12}^q/\Gamma_{12}^q)$ from the
mass difference $\dm_q\simeq 2|M_{12}^q| $ among the eigenstates, their
width difference $\dg_q \simeq 2\, |\Gamma_{12}^q| \cos \phi_q$ and the
semileptonic $CP$ asymmetry
\begin{eqnarray}
a^q_{\rm SL} &=& \imag \frac{\Gamma_{12}^q}{M_{12}^q} =
\frac{|\Gamma_{12}^q|}{|M_{12}^q|} \sin \phi_q \; = \;
\frac{\dg_q}{\dm_q} \tan \phi_q . \label{defafs}
\end{eqnarray}
Resulting from box diagrams with heavy (virtual) particles, $M_{12}^q$ is expected to be especially sensitive to NP~\cite{Lenz:2010gu}. Therefore the two complex parameters 
$\Delta_s$ and 
$\Delta_d$, defined as   
\begin{eqnarray}
M_{12}^q & \!\equiv\! & M_{12}^{\text{SM},q} \cdot  \Delta_q \, ,
\quad\;  \Delta_q  \equiv   |\Delta_q| e^{i \phi^\Delta_q} , 
\quad\; q=d,s, \; \label{defdel}
\end{eqnarray}
can differ substantially from the SM value $\Delta_s=\Delta_d=1$. 
Importantly, the NP phases $\phi^\Delta_{d,s}$ do not only
affect $a^{d,s}_{\rm SL}$, but also shift the $CP$ phases extracted from
the mixing-induced $CP$ asymmetries in $B_d \to J/\psi K$ and $B_s \to
J/\psi \phi$ to $2\beta+\phi^\Delta_d$ and $2\beta_s-\phi^\Delta_s$,
respectively. 
There has been a lot of interest triggered on this possibility due to disagreements with respect to the SM shown first by the early measurements
from CDF and D\O\ on the $B_s$ mixing angle, and further once D\O\ quoted values of the like-sign dimuon asymmetry $A_{\rm SL}$ (measuring a linear combination of $a^d_{\rm SL}$ and $a^s_{\rm SL}$). Later measurements of the individual semileptonic $CP$ asymmetries and mixing angles have not been able to explain the D\O\ measurement, as they showed a good agreement with SM expectations.

\begin{figure}[t]
  \includegraphics[width=8cm]{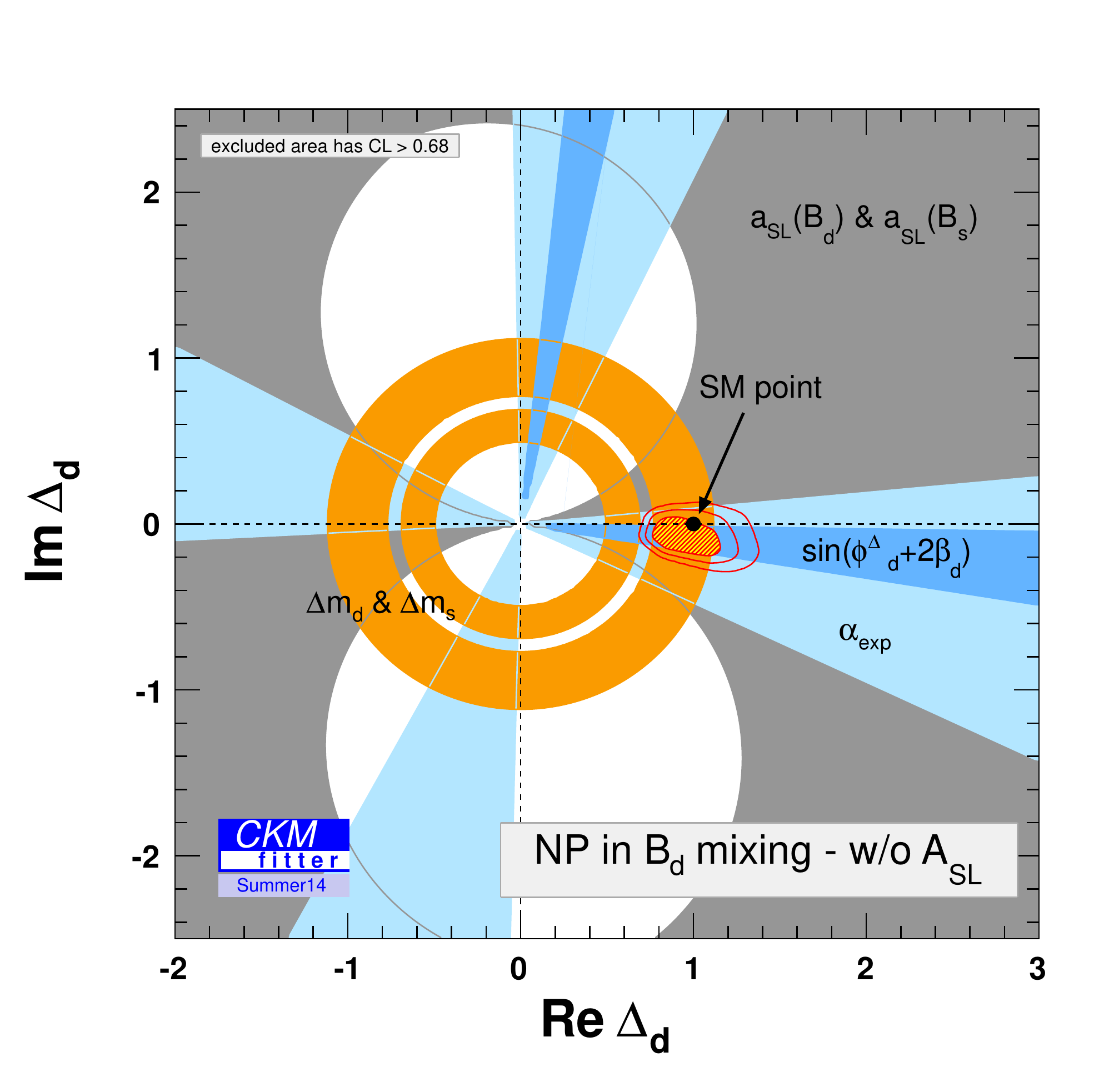}
   \includegraphics[width=8cm]{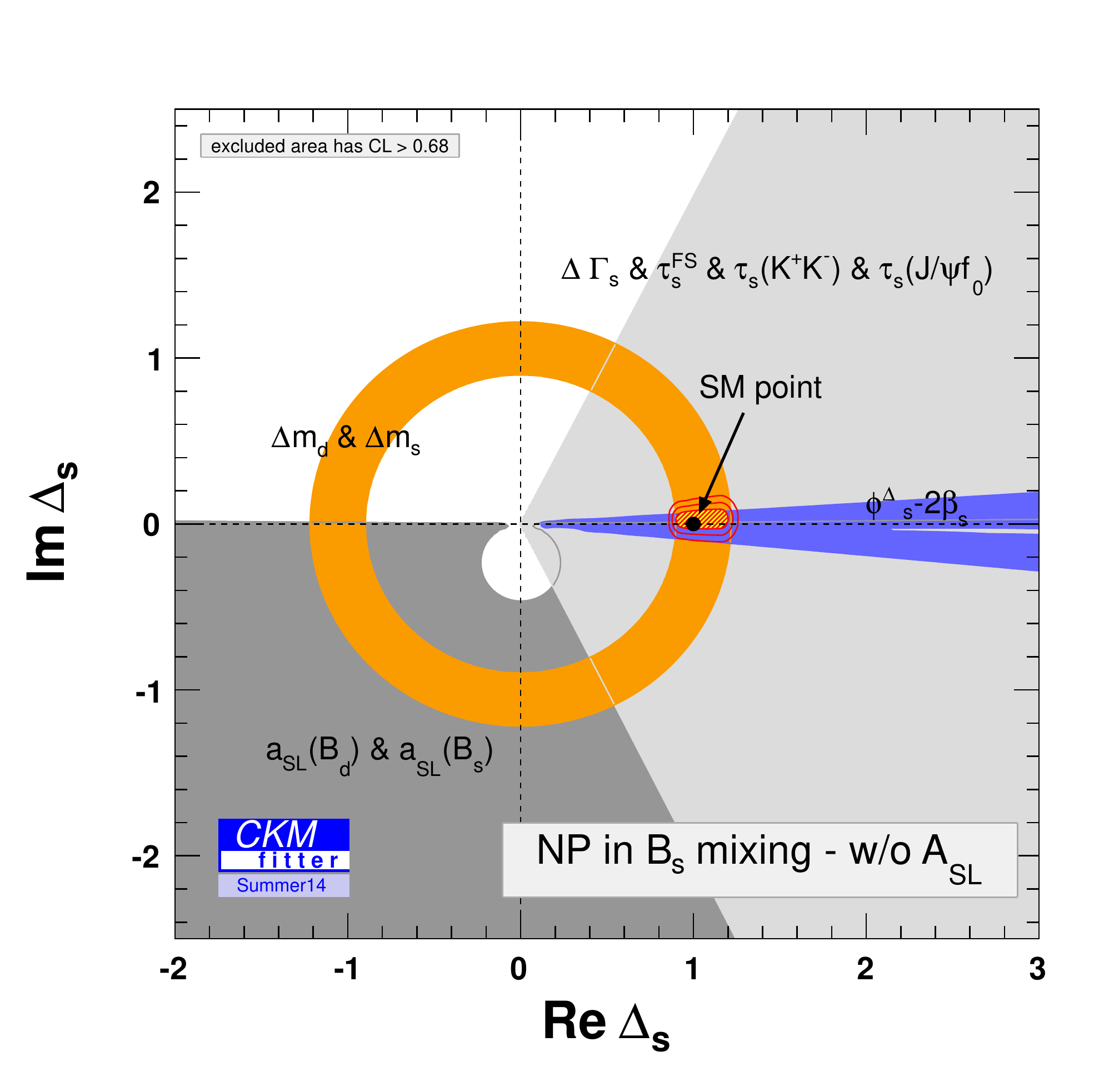}
\caption{\small Complex parameters  $\Delta_d$  (up) 
     and $\Delta_s$ (down) in Scenario I, not including $A_{\rm SL}$.  
          The coloured areas represent regions
     with $1-p < 68.3~\%$ for the individual constraints ($\alpha_{\rm exp}\equiv \alpha -\phi_d^\Delta/2$).
     The red area shows
     the region with $1-p < 68.3~\%$ for the combined fit, 
    with the two additional contours 
     delimiting the regions with $1-p < 95.45~\%$ and 
     $1-p < 99.73~\%$.   
     \label{fig:withoutASL}}
\end{figure}

\begin{figure}[t]
  \includegraphics[width=8cm]{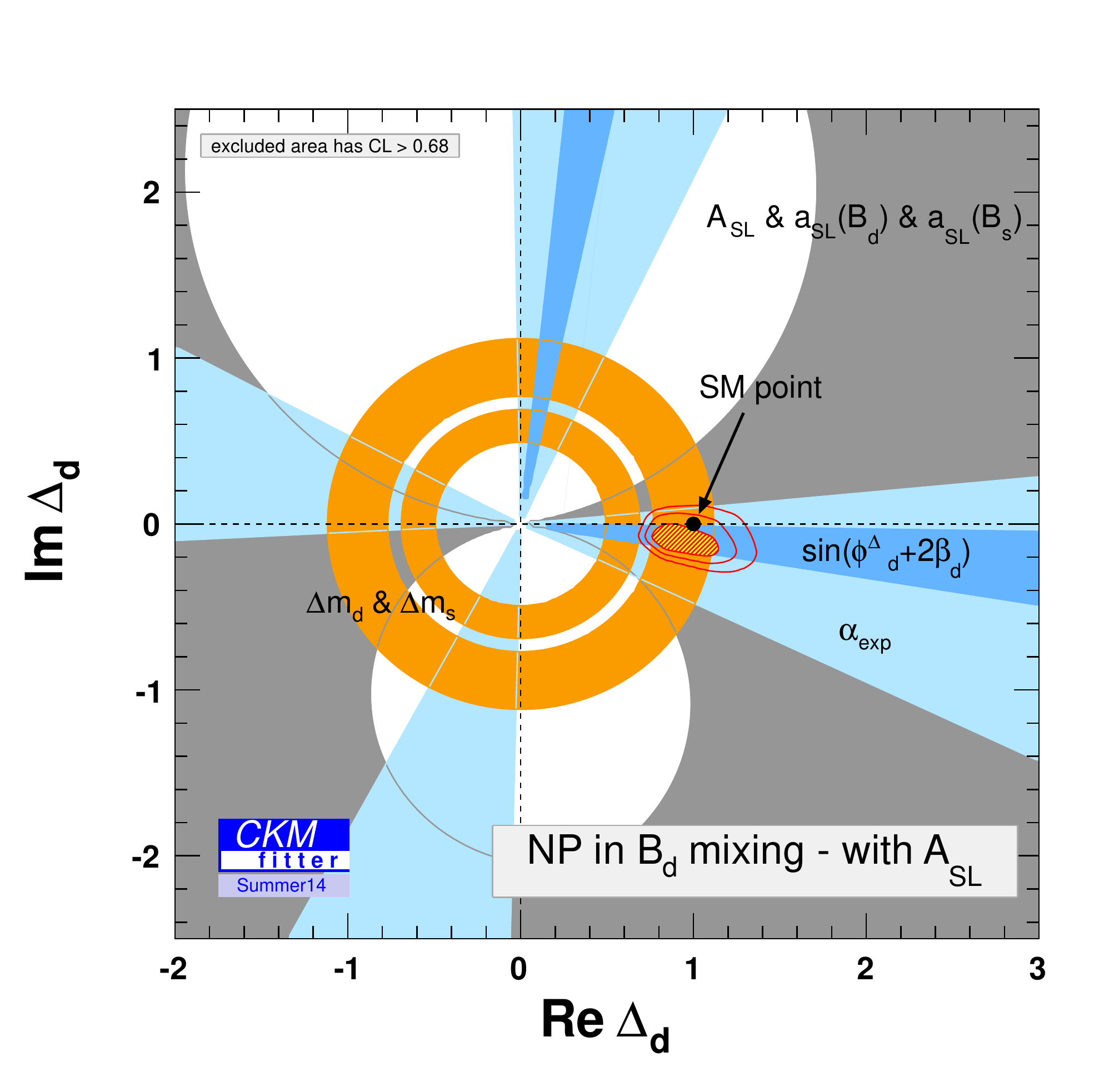}
   \includegraphics[width=8cm]{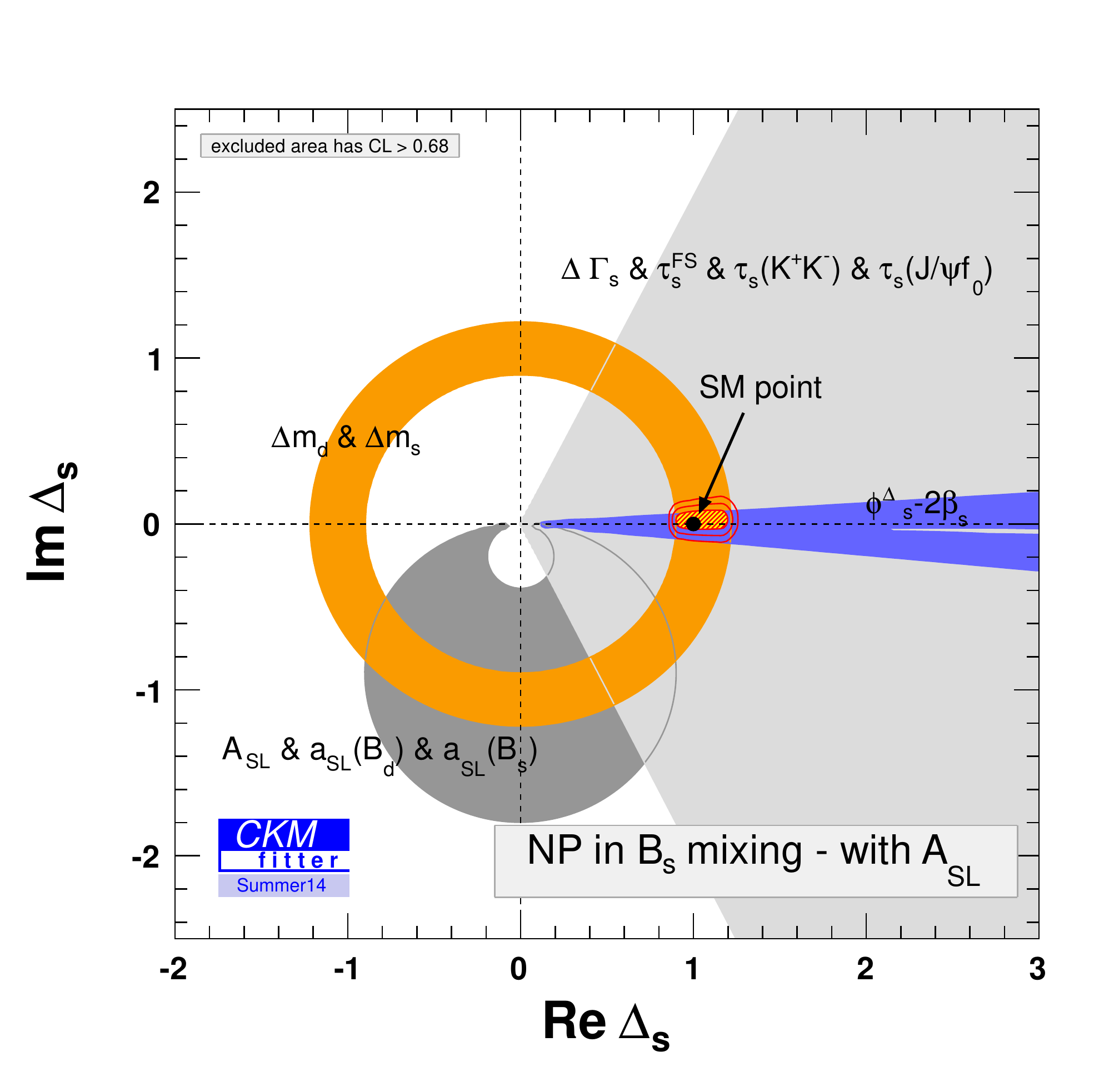}
\caption{\small Complex parameters  $\Delta_d$  (up) 
     and $\Delta_s$ (down) in Scenario I, including $A_{\rm SL}$. 
     The conventions are the same as in Fig.~\ref{fig:withoutASL}.
     \label{fig:withASL}}
\end{figure}

 In Refs.~\cite{Lenz:2010gu,Lenz:2012az} we have
determined the preferred ranges for $\Delta_s$ and $\Delta_d$ in a
simultaneous fit to the CKM parameters in different generic scenarios in
which NP is confined to $\Delta F=2$ flavour-changing neutral currents. 
We focus here on Scenario~I, where we have treated  $\Delta_s$ and $\Delta_d$  independently, corresponding to NP with arbitrary flavour structure. In this setting, $K\bar{K}$ involves three other, unrelated, new physics coefficients which will not be discussed in the following.
We use most of the inputs involved in the global fit, apart from ${\cal B}(B_s\to\mu\mu)$, which is likely to be also affected by New Physics in a way that cannot be connected simply to the New Physics introduced in $\Delta F=2$ processes. In Scenario~I, $\epsilon_K$ is affected by NP independently from the $B_d$ and $B_s$ sectors, and  thus has no impact on the discussion of NP here. The remaining parameters can be found in ref.~\cite{Lenz:2010gu,Lenz:2012az}.

One comment is in order concerning the recent reassessment of the value of $A_{\rm SL}$. Members of the $D\O$ experiment~\cite{Borissov:2013wwa} have considered an additional SM source for $CP$-violation contributing the dimuon charge asymmetry (coming from the interference of $b\to c\bar{c}s$ decay with and without mixing). This modifies the extraction of the linear combination of $a_{\rm SL}^d$ and $a_{\rm SL}^s$ from the like-sign dimuon asymmetry. This correction has been included in the latest $D\O$ update, bringing $A_{\rm SL}$ closer to its SM value. But the estimate of this correction has been challenged later~\cite{NiersteASL}, as it misses
other contributions from $b\to c\bar{u}s, u\bar{c}s, u\bar{u}s$ which could partially compensate this new correction. As the theoretical status remains unclear, and since $A_{\rm SL}$ has been in the past always very difficult to reconcile with the other $\Delta F=2$ observables even within our rather generic scenario, we will consider two sets of results, with and without the inclusion of the $D\O$ measurement.

In addition, we have updated the values of the bag parameters, following the recent work from the ETMC collaboration~\cite{Carrasco:2013zta}, working with $N_f=2$ dynamical flavours. The impact on our study is however small, since their  results showed an excellent compatibility with the previous (quenched) study~\cite{Becirevic:2001xt} that we used in previous publications. For the 
ratio of scalar quenched bag parameters, we have assumed that the breakdown between statistical and systematic errors in ref.~\cite{Carrasco:2013zta} followed the same pattern as for the SM ($B_1$) bag parameters.
All these additional inputs are collected in Table~\ref{tab:InputsNP}.

\begin{table}[t]
\renewcommand{\arraystretch}{1.2}
\begin{tabular}{lcc} 
Quantity   & without $A_{\rm SL}$ & with $A_{\rm SL}$ \\
 \hline  &&   \\[-0.3cm]
$\mbox{Re}{(\Delta_d)}$  &  $0.94^{+0.18}_{-0.15}$ & $0.88^{+0.22}_{-0.10}$ \\[0.15cm]
$\mbox{Im}{(\Delta_d)}$  &  $-0.12^{+0.12}_{-0.05}$ & $-0.11^{+0.07}_{-0.05}$ \\[0.15cm]
$|\Delta_d|$             &  $0.95^{+0.18}_{-0.15}$ & $0.89^{+0.22}_{-0.10}$\\[0.15cm]
$\phi^\Delta_d$ [deg]    &  $-6.9^{+6.9}_{-2.2}$ & $-7.3^{+4.7}_{-2.1}$\\[0.15cm]
$\mbox{Re}{(\Delta_s)}$  &   $1.05^{+0.14}_{-0.13}$ & $1.01^{+0.17}_{-0.09}$\\[0.15cm]
$\mbox{Im}{(\Delta_s)}$  &  $0.03^{+0.04}_{-0.04}$ & $0.02^{+0.04}_{-0.04}$   \\[0.15cm]
$|\Delta_s|$             &  $1.05^{+0.14}_{-0.13}$ & $1.01^{+0.17}_{-0.10}$ \\[0.15cm]
$\phi^\Delta_s$ [deg]    &$1.5^{+2.3}_{-2.4}$ & $1.3^{+2.3}_{-2.3}$\\[0.15cm]
$\phi^\Delta_d+2\beta$ [deg] (!)   &   $46.^{+13}_{-12}$ & $38^{+10}_{-13}$\\[0.15cm]
$\phi^\Delta_s-2\beta_s$ [deg] (!) &  $ -49.^{+43}_{-16}$   & $-61^{+13}_{-5}$  \\[0.15cm]
\hline &&      \\[-0.3cm]
$A_\text{SL}$~~$[10^{-4}]$ (!)    & $-7.1^{+3.7}_{-4.3}$ & $-7.1^{+3.7}_{-4.3}$ \\[0.15cm]
$A_\text{SL}$~~$[10^{-4}]$        & $-$ & $-10.4^{+4.7}_{-2.2}$ \\[0.15cm]
$a_\text{SL}^{d}$~~$[10^{-4}]$ (!)            
                &  $-17.3^{+7.6}_{-5.9}$ or $121^{+35}_{-43}$  & $-20.7^{+6.8}_{-3.8}$ \\[0.15cm]
$a_\text{SL}^{s}$~~$[10^{-4}]$ (!)            
                & $1.6^{+1.9}_{-1.9}$ & $1.5^{+1.9}_{-1.9}$ \\[0.15cm]
$\Delta\Gamma_d [\mathrm{ps}^{-1}]$  (!)            
                &  $0.0028^{+0.0018}_{-0.0006}$  & $0.0042^{+0.0005}_{-0.0019}$ \\[0.15cm]
$\Delta\Gamma_s [\mathrm{ps}^{-1}]$ (!)            
                & $0.090^{+0.082}_{-0.024}$ &  $0.089^{+0.082}_{-0.023}$\\[0.15cm]
$\Delta\Gamma_s [\mathrm{ps}^{-1}]$            
                & $0.081^{+0.006}_{-0.006}$ &$0.081^{+0.006}_{-0.006}$  \\[0.15cm]
\hline &&      \\[-0.3cm]
$B\to \tau\nu$~~$[10^{-4}]$ (!) & $0.688^{+0.380}_{-0.048}$ & $1.033^{+0.065}_{-0.345}$  \\[0.15cm]
$B\to \tau\nu$~~$[10^{-4}]$     & $1.029^{+0.062}_{-0.201}$ & $1.037^{+0.062}_{-0.155}$ \\[0.15cm]
\end{tabular}

\caption{68.3\% CL intervals for the results of the fits in Scenario I, including or not the $A_{\rm SL}$ measurement. The notation (!) means that the fit output represents the indirect constraint with the corresponding direct input  removed.\label{tab-results}}
\end{table}

\begin{table}[t]
\begin{center}
\begin{tabular}{l|cc|cc}
 & \multicolumn{2}{c|}{Without $A_{\rm SL}$} & \multicolumn{2}{c}{With $A_{\rm SL}$}\\
Quantity(ies) &  \multicolumn{2}{c|}{Deviation wrt} &   \multicolumn{2}{c}{Deviation wrt} \\ 
& SM & Sc. I  & SM & Sc. I  \\
 \hline \\[-0.3cm]
$\phi^\Delta_d+2\beta$ & $1.6~\sigma$ & $0.0~\sigma$ & $1.6~\sigma$ & $0.0~\sigma$\\[0.15cm]
$\phi^\Delta_s-2\beta_s$ & $0.0~\sigma$ & $1.1~\sigma$ & $0.0~\sigma$ & $2.6~\sigma$\\[0.15cm]
\hline &      \\[-0.3cm]
$A_{\rm SL}$ & $-$ & $-$ & $2.7~\sigma$ & $2.4~\sigma$  \\[0.15cm]
$a_{\rm SL}^{d}$ & $0.4~\sigma$ & $0.8~\sigma$ & $0.4~\sigma$ & $1.1~\sigma$  \\[0.15cm]
$a_{\rm SL}^{s}$ & $1.0~\sigma$ & $1.0~\sigma$ & $1.0~\sigma$ & $1.0~\sigma$  \\[0.15cm]
$\Delta\Gamma_s$ & $0.3~\sigma$ & $0.3~\sigma$ & $0.1~\sigma$ & $0.1~\sigma$   \\[0.15cm]
\hline  \\[-0.3cm]
$\mathcal{B}(B\to\tau\nu)$ & $1.3~\sigma$ & $0.8~\sigma$ &  $1.3~\sigma$ & $0.2~\sigma$  \\[0.15cm]
\hline  \\[-0.3cm]
$\mathcal{B}(B\to\tau\nu)$, $A_{\rm SL}$ & $-$ & $-$ &  $2.5~\sigma$ & $2.1~\sigma$  \\[0.15cm]
$\phi_s^\Delta-2\beta_s$, $A_{\rm SL}$ & $-$ & $-$ & $2.2~\sigma$& $2.2~\sigma$ \\[0.15cm]
$\mathcal{B}(B\to\tau\nu)$, $\phi_s^\Delta-2\beta_s$, $A_{\rm SL}$ & $-$ & $-$ & $2.2~\sigma$ &  $1.9~\sigma$ 
 \end{tabular} 
\end{center}
\caption{Pull values for selected parameters and observables in SM and Scenarios I in terms of the number of 
equivalent standard deviations between the direct measurement and the full indirect
fit predictions. Two different types of fits, including or not $A_{\rm SL}$ are considered.}\label{tab-pulls}
\end{table}

\subsection{Constraints on New Physics}

We summarise our results  in
Tables~\ref{tab-results} and \ref{tab-pulls} and in
Figs.~\ref{fig:withASL} and \ref{fig:withoutASL}, including or not $A_{\rm SL}$.
We find pull values for $A_{\rm SL}$ and $\phi_s^\Delta-2\beta_s$ of
 2.4$\,\sigma$ and 2.5$\,\sigma$ respectively, illustrating the discrepancy between the two
 constraints in Fig.~\ref{fig:withASL}.  We do not
quote pull values for $\Delta m_{d,s}$ in Sc.~I, as these observables
are not constrained
once their experimental measurement is removed. 

The global constraint on the argument of $\Delta_s$ is more stringent than what could be assumed by the overlap of the constraints from $\Delta m_d$, $\Delta m_s$ and $\phi_s^\Delta-2\beta_s$. This can be understood as follows: the fit including NP in $\Delta F=2$ has a discrete ambiguity in the determination of $\bar\rho,\bar\eta$, so that two solutions (symmetrical with respect to the origin) are allowed~\cite{Lenz:2010gu,Charles:2013aka}. This translates into two possibilities for $\beta_{sb}$, with opposite signs. The constraint from $\phi_s^\Delta-2\beta_s$ also exhibits two preferred solutions for $\arg(\Delta_s)$. These two solutions cannot be distinguished at 1 $\sigma$ if only $\phi_s^\Delta-2\beta_s$ is considered, but the degeneracy is lifted in favour of the SM-like solutions once the other constraints are added, leading to a global constraint centered around the solution corresponding to the SM-like solution for $\bar\rho,\bar\eta$, with a domain smaller in size than the constraint
from $\phi_s^\Delta-2\beta_s$

The comparison between the fits with and without $A_{\rm SL}$ shows a slight decrease for $|\Delta_d|$ when $A_{\rm SL}$ is added, whereas
$|\Delta_s|$ is essentially unchanged. One notices also that in the absence of $A_{\rm SL}$, the predicted value of $a_\text{SL}^{d}$
can take two different values (a small negative one or a large positive one), corresponding to the two 
branches allowed by $\phi_d^\Delta+2\beta$. The predicted value for $\phi_d^\Delta+2\beta$ varies significantly
when $A_{\rm SL}$ is added or not, since it comes from the combination of the constraint from $\alpha$ measurements with the semileptonic asymmetries. This yields a 
noticeable change in the prediction for $\Delta \Gamma_d$. Even though the predictions for $\Delta \Gamma_s$ and $B\to\tau\nu$ also seem  to vary, this  mainly concerns the best-fit point and is much less the case once $1\sigma$ intervals are considered.

One can also consider the $p$-value of the SM hypothesis following the discussion in ref.~\cite{Lenz:2010gu}.
Without $A_{\rm SL}$, the $p$-value for the 2D SM hypothesis $\Delta_d=1$ ($\Delta_s=1$) is
     0.9~$\sigma$ (0.3~$\sigma$), and the 4D SM hypothesis $\Delta_d=1=\Delta_s=1$
     has a p-value of 0.7~$\sigma$.
With $A_{\rm SL}$, the $p$-value for the 2D SM hypothesis $\Delta_d=1$ ($\Delta_s=1$) is
     1.2~$\sigma$ (0.3~$\sigma$), and the 4D SM hypothesis $\Delta_d=1=\Delta_s=1$
     has a p-value of 1.0~$\sigma$.
    
     The two complex NP
parameters $\Delta_d$ and $\Delta_s$ (parametrising NP in
$M_{12}^{d,s}$) are not sufficient to absorb the discrepancy between the D\O\ measurement of 
$A_{\rm SL}$ and the rest of the global fit. The situation
has however improved compared to earlier analyses, due to the decreased discrepancy
of $A_{\rm SL}$  compared to the Standard Model. Without $A_{\rm SL}$, the fit including NP in $\Delta F=2$ is good, but 
the improvement with respect to the SM is limited.
In addition, we stress that data still allow sizeable NP contributions  in both $B_d$ and $B_s$ sectors up to
30-40\% at the 3$\sigma$ level.

\section{Conclusion}

This article collects a selection of SM predictions
driven by the global fit of the CKM parameters, in view of related
recent or foreseeable experimental measurements.   
The results were obtained by combining the inputs  collected
in Table~\ref{tab:expinputs}, using the statistical frequentist framework adopted by the CKMfitter group.
The overall agreement of the Standard Model global fit is impressive, as confirmed by the representation of the various unitarity triangles and the results given in Table~\ref{tab:pred:meas}, gathering the SM predictions using the inputs. We discussed the status of some quantities of importance whose status has changed recently ($\alpha$, $\gamma$, ${\cal B}(B_s\to\mu\mu)$). We also provided predictions for various observables of interest, as well as a  table of pulls. 

We have also performed a global fit to flavour physics data 
in a scenario with generic New Phiscs in the $B_d\bar{B}_d$ and $B_s\bar{B}_s$ amplitudes,
as defined in Refs.~\cite{Lenz:2010gu,Lenz:2012az}. 
The discrepancy between $A_{\rm SL}$ and the rest of the neutral-meson mixing observables remains even in this extended scenario. If we remove $A_{\rm SL}$, because of the potentially large (and unknown) systematics affecting
its extraction, the fit improves significantly, with a SM-like scenario being very likely. However, significant contributions from NP are still possible at the 3$\sigma$ level. This is an invitation for more study of these observables with the LHCb upgrade and the start of Belle-II, as discussed in the prospective exercise of 
ref.~\cite{Charles:2013aka}. 

\vspace{0.2cm}

\acknowledgments

L.V.S. acknowledges financial support from the Labex P2IO (Physique des 2 Infinis et Origines). P.U acknowledges financial support from the Australia Research Council under the Future Fellowship program (FT130100303).

\end{document}